\documentclass[prb,twocolumn,superscriptaddress]{revtex4}

\usepackage{amsmath} 
 
\usepackage{graphicx} 
\usepackage{times}

\begin{document}

 
\title{Edge states and tunneling of non-Abelian quasiparticles
in the $\nu=5/2$ quantum Hall state and $p+ip$ superconductors}

\author{Paul Fendley} 
\affiliation{Department of Physics, University of Virginia, 
Charlottesville, VA 22904-4714} 
\author{Matthew P.A. Fisher} 
 
\affiliation{Kavli Institute for Theoretical Physics, 
University of California, Santa Barbara, CA 93106-4030} 
\author{Chetan Nayak} 
\affiliation{Microsoft Research, Project Q, Kohn Hall, 
University of California, Santa Barbara, CA 93106-4030} 
\affiliation{Department of Physics and Astronomy, 
University of California, Los Angeles, CA 90095-1547}

\date{July 14, 2006}

\begin{abstract}
We study quasiparticle
tunneling between the edges of a non-Abelian topological state.
The simplest examples are
a $p+ip$ superconductor and the Moore-Read Pfaffian
non-Abelian fractional quantum Hall state; the latter
state may have been observed at Landau level filling
fraction $\nu=5/2$.
Formulating the problem
is conceptually and technically non-trivial:
edge quasiparticle correlation functions are 
elements of a vector space, and
transform into
each other as the quasiparticle coordinates
are braided. We show in general how to resolve this difficulty and
uniquely define the quasiparticle tunneling Hamiltonian.
The tunneling operators in the simplest examples
can then be rewritten in terms of a free boson.
One key consequence of this bosonization
is an emergent spin-$1/2$ degree of freedom. We show that
vortex tunneling across a $p+ip$ superconductor
is equivalent to the single-channel Kondo problem, while quasiparticle
tunneling across the Moore-Read state
is analogous to the two-channel Kondo effect.
Temperature and voltage dependences of the tunneling
conductivity are given in the low- and high-temperature limits.
\end{abstract}

\maketitle 
 

\section{Introduction}

The possible existence of non-Abelian quantum Hall states
has caused great excitement recently \cite{Physics-today,Sci-Am}.
A non-Abelian quantum Hall state would not only be
a new class of quantum matter -- a truly remarkable
discovery in itself -- but could also be a platform for
fault-tolerant quantum computation \cite{tqc}. The leading candidate
is the observed \cite{Willet87,Eisenstein02,Xia04} quantized Hall plateau
with $\sigma_{xy}=\frac{5}{2}\,\frac{e^2}{h}$.
There is numerical evidence \cite{Morf98,Rezayi00} that
the ground state at this filling fraction is
given by a filled lowest Landau level of both spins
and $\nu=1/2$ filling of the first excited Landau level in
the Moore-Read Pfaffian state \cite{Moore91}.
The excitations of this state are charge-$e/4$ quasiparticles
which exhibit non-Abelian braiding statistics
\cite{Nayak96c,Read96,Fradkin98,Tserkovnyak03}.

The Moore-Read Pfaffian state is the quantum Hall incarnation
of a $p+ip$ superconductor \cite{Greiter92}, whose
vortices have the same non-Abelian braiding statistics
\cite{Read00,Ivanov01,Stern04,Stone06}. Such vortices
have Majorana (real) fermion zero modes in their cores. A pair
of vortices, if kept far apart,
therefore shares a complex fermion zero mode which can
be either occupied or unoccupied. As vortices are
braided, the occupancies of these zero modes
are altered and phases are acquired. Such transformations
do not commute, so the braiding statistics of vortices
is non-Abelian. There are at least
two candidate systems in which a $p+ip$ superconducting
state may exist: (1) the seemingly unconventional superconductor
Sr$_2$RuO$_4$ \cite{Mackenzie03} and
(2) ultra-cold fermions with a $p$-wave
Feschbach resonance in an atomic trap \cite{Gurarie05}.
These systems may be fertile alternatives to
the $\sigma_{xy}=\frac{5}{2}\,\frac{e^2}{h}$ quantum
Hall state for exploring non-Abelian braiding statistics and
topological quantum computation \cite{DasSarma06,Tewari06}.

Multi-quasiparticle states of non-Abelian anyons are topologically
degenerate. These states can be used as decoherence-free qubits
which can be manipulated and measured using
quasiparticle braiding and interferometry.
This observation is the basis not only of proposals
for quantum computation \cite{DasSarma05,Bravyi05,Freedman06}
but also for experiments to determine
whether the $\sigma_{xy}=\frac{5}{2}\,\frac{e^2}{h}$ state
is non-Abelian \cite{Fradkin98,DasSarma05,Stern05,Bonderson05,Hou06} .
All of these proposed experiments involve interference
between trajectories in which non-Abelian quasiparticles
tunnel at point contacts between the edges of a Hall droplet.

A better understanding of the behavior of a point contact
would illuminate the analysis of these proposed experiments.
It would also open a new avenue for exploring
the physics of the $\sigma_{xy}=\frac{5}{2}\,\frac{e^2}{h}$ state
-- and, perhaps, Sr$_2$RuO$_4$ and
ultra-cold fermions with a $p$-wave Feschbach resonance --
because transport through a point contact is an important probe
of quantum Hall states. Chiral topological states have
gapless edge excitations whose behavior is largely
determined by the topological properties of the bulk \cite{Wen91}.
These gapless excitations determine low-temperature
transport properties. At a point contact, fractionally-charged
quasiparticles tunnel from one edge of the system to the other.
Consequently, the temperature and voltage dependences
for transport through a point contact reflect the topological
structure of the state. In the case of the Laughlin states,
a small bare tunneling rate between
the two sides of a Hall bar at any finite temperature
increases as the temperature
is decreased until the bar is effectively broken in two
at zero temperature.
The conductance versus temperature power laws in
both the high-temperature and low-temperature limits
\cite{Kane92} (and even the full crossover function between
these two limits \cite{Fendley95}) show the effects of the
fractional statistics of quasiparticles in these states.
Shot noise and other measurements
evince the fractional charge of quasiparticles \cite{Reznikov97}. In the case
of the hierarchy states, the topological structure is richer,
but has proven more elusive experimentally \cite{Grayson98}.
One might also expect even more interesting physics in a single point contact
in a non-Abelian quantum Hall state \cite{Bena06}, reflecting
its topological properties.

In ref. \onlinecite{Fendley06a}, we analyzed the behavior of
a single point contact in the Moore-Read Pfaffian quantum
Hall state and also in the slightly simpler case of
a $p+ip$ superconductor and found that it is highly non-trivial.
A point contact between two edges of a
$\nu=5/2$ Moore-Read Hall droplet leads to a leading
correction to the vanishing of the longitudinal resistivity
$R_{xx}\sim T^{-3/2}$ (at temperatures which are sufficiently
high that this is a small correction). At zero temperature,
the filled lowest Landau level is unaffected but the
$\nu=1/2$ first excited Landau level is broken in two
so that $R_{xx} = \frac{1}{10}\,\frac{h}{e^2}$ (see appendix
\ref{sec:four-terminal} for the definition of the four-terminal
resistance). At small non-zero temperature,
$R_{xx} - \frac{1}{10}\,\frac{h}{e^2} = - {T^4}$.
We showed that the crossover between these two limits is
a variant of the two-channel Kondo problem
\cite{Emery92} and resonant tunneling in Luttinger liquids \cite{Kane92}.
In this paper, we explain in greater detail how to properly
define tunneling at a point contact in a
non-Abelian state.  We expand upon our construction
of a bosonized representation
for the tunneling Hamiltonian, thereby clarifying
the relation to the Kondo problem.
As in the Kondo problem, the crossover
from high to low-temperatures
is accompanied by entropy loss.
In a companion paper\cite{Fendley06c}, we show that there is a very
natural $2+1$-dimensional interpretation for this
entropy loss.

The edge excitations of the Laughlin and hierarchy states
correspond to free chiral bosons. Although these are
free field theories, the perturbation corresponding to
a point contact, at which quasiparticles tunnel
between edges, is built up from exponentials of
the chiral boson. Such operators are
non-trivial and capture the fractional charge and
Abelian fractional statistics of quasiparticles.
The Moore-Read Pfaffian state has a further
wrinkle: in addition to a chiral boson there is a Majorana
fermion edge mode \cite{Milovanovic96}. Although this, too,
is a free field theory it is a more peculiar one than a
chiral boson. The Majorana fermion is the chiral part
of the critical theory of the $2D$ Ising model, which has
non-trivial spin-spin correlation functions.
This is related, as we will see below,
to the non-Abelian statistics of the quasiparticles.
The edge of a $p+ip$ superconductor
has only this Majorana mode -- it lacks a charge-carrying
chiral boson mode -- but the same issues arise. 

The non-Abelian statistics of the state plays a crucial role in
describing quasiparticle tunneling at a point contact. The operator
which creates a charge $e/4$ quasiparticle at the edge of the
Moore-Read Pfaffian state at $\nu=5/2$ or a vortex at the edge of a
$p+ip$ superconductor is the chiral part of the Ising spin field
$\sigma({\bf r})$, which creates a branch cut for the fermions, terminating
at ${\bf r}$. Correlation functions of the tunneling
operator therefore involve the chiral parts of spin-spin correlation
functions. The chiral parts of correlation functions in a conformal
field theory, such as the Ising model, are called {\it conformal
blocks} \cite{Belavin84}. In general, they are not defined uniquely if
only the positions of the fields are specified. As these coordinates
are taken around each other (i.e.\ {\em braided}), the conformal
blocks transform linearly. In other words, the conformal blocks form a
vector space, on which the braid group is represented. In the
case of exponentials of a chiral boson, braiding simply results in a
phase. In the fractional quantum Hall context, this is the
Abelian braiding statistics of quasiparticles in the Laughlin and
hierarchy states. However, in a generic rational conformal field
theory \cite{Moore88}, there is a multi-dimensional space of such
chiral parts of correlation functions, i.e.\  of conformal blocks. 
The full non-chiral correlation function is a sum of products of left- and
right-conformal blocks, and must be single-valued. The consequent
constraints on the conformal blocks result in a great deal of
structure, which is discussed in depth in Refs.\
\onlinecite{Moore88}.

In order to define the tunneling Hamiltonian for non-Abelian
quasiparticles and compute the effects of a point contact
perturbatively, we must compute multi-point chiral correlation
functions such as $\left\langle {\sigma} {\sigma} \ldots
{\sigma}\right\rangle$. In conformal field theory language, the
ambiguity in defining such a quantity stems from the two possible
fusion channels for a pair of spin fields, written schematically as
$\sigma \cdot\sigma \sim 1 + \psi$.  Each of these possibilities (or
any linear combination thereof) corresponds to a different possible
chiral correlation function.  Consequently, the vector space
of chiral correlation functions with $2n$ spins is
$2^{n-1}$-dimensional.  This can be restated in different terms by
observing that a pair of quasiparticles can be in either of two
topologically distinct states \cite{Nayak96c,Read96,Read00}, the two
states of the qubit which they form \cite{DasSarma05}.  In the
language of a $p+ip$ superconductor, the complex fermion zero mode
associated with a pair of vortices can be either occupied or
unoccupied -- the two states $|0\rangle$ and $|1\rangle$ of the qubit.
When a pair of quasiparticles is in the state $|0\rangle$, they fuse
to the $1$; when they are in the state $|1\rangle$, they fuse to
$\psi$. Thus, in order to properly define these correlation functions,
we must also specify the state of the qubit associated with each pair
of quasiparticles. How we pair them is arbitrary; changing the pairing is
just a change of basis.

The resolution of this problem can be stated quite simply in
physical terms: when a quasiparticle tunnels, a neutral fermion
cannot be created by the tunneling process alone. Therefore,
the two quasiparticle operators (corresponding to the
annihilation of a quasiparticle on one edge and its subsequent
creation on the other edge) must fuse to the identity.
However, this is not the most convenient basis for
computations. We would rather know how quasiparticle
operators on the same edge fuse so that different edges
can be treated perturbatively as being independent.
This is simply a basis change in the space of conformal blocks.
We can switch into such a basis using the braiding rules
of the chiral Ising conformal field theory \cite{Moore88}
or, equivalently, the corresponding topological field theory \cite{Witten89}. 

The basic strategy outlined above can be applied to any
two-dimensional gapped system with gapless edge modes described by
conformal field theory. However, in the case of the Moore-Read
Pfaffian state, we can massage this result into an even simpler
form. When considering tunneling between two different edges, we
combine the Majorana fermion modes at the two edges into a single
Dirac fermion. We then bosonize this Dirac fermion.  This allows us to
directly compute the conformal blocks of the critical $2D$ Ising
model. In order to bosonize the quasiparticle tunneling operator, we
need to introduce a spin-$1/2$ degree of freedom. Although we
introduce this degree of freedom almost as a bookkeeping device, we
then find that the bosonized tunneling Hamiltonian takes a form
similar to the anisotropic Kondo Hamiltonian. Armed with this
knowledge, we analyze the crossover to the low-temperature limit in
which tunneling becomes strong.

Sections \ref{sec:vortices}, \ref{sec:Ising}, and \ref{sec:edgemodes}
are reviews in which
we explain why the edge excitations
of the Moore-Read Pfaffian state and of a $p+ip$ superconductor
contain a Majorana fermion mode which is the chiral
part of the critical $2D$ Ising model. 
In section \ref{sec:point-contact},
we discuss the form of the tunneling Hamiltonian. We explain
in detail the subtlety associated with defining this Hamiltonian
and show how to resolve this difficulty in sections
\ref{sec:conformal-blocks} and \ref{sec:decomposition}. We then show
in section \ref{sec:bosonization} how this Hamiltonian can be bosonized.
In section \ref{sec:Kondo}, we map the bosonized Hamiltonian
to the Kondo and Luttinger liquid resonant tunneling Hamiltonians.
In section \ref{sec:instanton}, we analyze the infrared behavior
of this bosonic Hamiltonian using these mappings and
an instanton gas expansion.

\section{Vortices and Edge States in a $p_x + i p_y$ Superconductor}
\label{sec:vortices}

The physics of a $p_x + i p_y$ superconductor is essentially identical
to the neutral sector of the Moore-Read Pfaffian
state \cite{Read00,Ivanov01,Stern04,Stone06}.
The vortices in the superconductor correspond to the
non-Abelian quasiparticles in the Moore-Read Pfaffian state, and the 
superconductor has a gapless chiral Majorana fermion edge mode
which is identical to the neutral sector of the Moore-Read Pfaffian
edge theory, which we will discuss in section \ref{sec:edgemodes}. 
Moreover, one can consider the process of passing a bulk vortex off
the edge of the superconductor, which leaves behind a ``twist'' field
or ``spin'' field operator $\sigma$ (a terminology which we
explain in section \ref{sec:Ising}) acting on the chiral edge state. 
This spin field changes the boundary conditions on the
chiral edge state from periodic to anti-periodic, or vice versa.
In this section we study the $p+ip$ superconductor at the level
of the Bogoliubov-De Gennes equations, to
gain insight into the physics of edge states and quasiparticles
in this non-Abelian topological state.

The Bogoliubov Hamiltonian for the $p+ip$ superconductor is expressed in
terms of the (spinless) fermion creation and destruction operators,
$\hat{c}(\bf{x}),\hat{c}^\dagger(\bf{x})$, with ${\bf x}$ denoting a
two-dimensional spatial coordinate. 
We define the Pauli matrices ${\vec{\sigma}}$ to act on the
two-component spinor
$$
\hat{\Psi}(\bf{x})  = \begin{pmatrix}
\hat{c}^\dagger(\bf{x})\\  \hat{c}\,(\bf{x}) 
\end{pmatrix}.
$$ 
The appropriate Bogoliubov-deGennes Hamiltonian is
\begin{equation}
\hat{\cal H} = \int d{\bf x}\, \hat{\Psi}^\dagger H \hat{\Psi}  ,
\end{equation}
with single-particle Hamiltonian,
$$
H = (-\nabla^2/2m + V({\bf x}) - \mu ) \sigma^z + i \Delta (\sigma^x \partial_x + \sigma^y
\partial_y )   .
$$
This  Hamiltonian has the symmetry, $\sigma^x H^* \sigma^x = - H$,
which implies that all non-zero-energy eigenstates come in $\pm E$ pairs.
To wit, with $H \phi_E = E \phi_E$ for a two-component wave function $\phi_E$,
one has $H \phi_{-E} = -E \phi_{-E}$ for $\phi_{-E} = \sigma^x \phi_E^*$.  One can then
expand the spinor field operator as
\begin{equation}
\label{eqn:bulk-modes}
\hat{\Psi} = \frac{1}{\sqrt{2}} \sum_{E > 0} [ \hat{\eta}_{E}  \phi_{E} +
\hat{\eta}_{-E} \phi_{-E}] ,
\end{equation}
with fermion operators $\hat{\eta}_{E}$ which satisfy $\hat{\eta}_{-E}^\dagger = \hat{\eta}_E$. 
The Hamiltonian can then be written in a diagonal form,
\begin{equation} 
\hat{\cal H} = \sum_{E>0} E \, \hat{\eta}_{E}^\dagger \hat{\eta}_{E}   .
\end{equation}
The ground state consists of filling up all of the negative energy states,
$|\text{Ground}\rangle = \prod_{E>0} \hat{\eta}_{-E}^\dagger | \text{vac}\rangle$, and is annihilated by
$\hat{\eta}_{E}$ and $\hat{\eta}^\dagger_{-E}$ for all $E > 0$. 
In the bulk of the superconductor all states will be gapped, obeying
$|E| >p_F \Delta $, with $p_F$ the Fermi momentum.

In order to establish the presence of chiral edge modes, it is convenient to
consider an infinite system
in which the potential $V(\bf{\bf{x}})$ varies spatially,
\begin{equation}
V(x,y) - \mu = \Delta \,V_0(y) ,
\end{equation} 
with $V_0(y>0)$ positive and increasing to large values for large y,
and $V_0(y<0)$ negative.  Since the electron density will fall to zero
for large positive $y$, we have, in effect, created a straight edge at
$y=0$.  At low energy, we can ignore the first term in the
single-particle Hamiltonian, $H$, because its second derivative makes
it smaller than the zero and single-derivative terms. For this
potential, there are exact eigenstates which are spatially
localized near $y=0$:
\begin{equation}
\label{eqn:zero-mode-wvfn}
\phi^{edge}_E({\bf x}) = e^{ik  x} e^{- \int_0^y V_0(y^\prime) dy^\prime}
\phi_{0} ,
\end{equation}
with $\phi_{0} = {1\choose 1}$ an eigenstate of $\sigma^x$.  This
wavefunction describes a chiral wave propagating in the $x-$direction
localized on the edge, with wave vector $k = E/\Delta$.  One can then
construct a second quantized description of these edge modes by
expanding the spinor field operator in terms of both bulk states above
the gap and an edge sector:
\begin{equation}
\hat{\Psi}({\bf x}) = \hat{\Psi}_{bulk}({\bf x}) + \hat{\Psi}_{edge}(\bf{x}) ,
\end{equation}  
with $\hat{\Psi}_{bulk}$ given in Eq. \ref{eqn:bulk-modes}, and
$$ \hat{\Psi}_{edge}({\bf x}) = e^{-\int_0^y V_0(y^\prime) dy^\prime}
\sum_{k >0} [ \hat{\psi}_k e^{ikx} \phi_0 + \hat{\psi}_{-k}
e^{-ikx} \sigma^x \phi_0^* ].
$$
Here the fermion operators, $\hat{\psi}_{\pm k}$,  satisfy $\hat{\psi}_{-k} = \hat{\psi}_k^\dagger$.
One sees that $\hat{\psi}(x) = \sum_k \hat{\psi}_k e^{ikx}$ is a real
Majorana field, $\hat{\psi}(x) = \hat{\psi}^\dagger(x)$ satisfying anticommutation
relations,
\begin{equation}
\left\{ \hat{\psi}(x),\hat{\psi}(x^\prime)\right\} = 2 \delta(x-x^\prime)  .
\end{equation}
The edge Hamitonian
can be simply written in terms of a real-space
Hamiltonian density:
\begin{equation}
\label{eqn:BdG-edge-Ham}
\hat{\cal H}_{edge}
= \sum_{k>0} v_n k \,\hat{\psi}_k^\dagger \hat{\psi}_k   
= \int dx \,\hat{\psi}(x) (-iv_n \partial_x ) \hat{\psi}(x)  ,
\end{equation}
where the edge velocity $v_n \equiv \Delta$. The Lagrangian density describes 
chiral Majorana edge modes propagating at velocity $v_n$,
\begin{equation}
{\cal L} =  i {\psi}(x) (\partial_t + v_n \partial_x) {\psi}(x).
\end{equation}
This is identical to Eq.~(\ref{eqn:L-neutral}) which will arise in
section \ref{sec:edgemodes} in our description of the neutral edge sector
of the Moore-Read state.

Next consider the edge of a large circular sample, with circumference
$L$.  Locally, each part of the edge looks ``flat", and one expects
that the edge wavefunction should take the same form as in
Eq. \ref{eqn:zero-mode-wvfn}.  The main difference is that the spinor,
$\phi_0$ in Eq.\ref{eqn:zero-mode-wvfn}, which describes a spin
one-half pointing in the $x-$direction, will have a direction in spin
space which is parallel to the local edge tangent vector.  Then, upon
fully encircling the drop, the spinor will rotate by $2\pi$ around the
$z-$axis, and will change sign.  This implies that in the edge
Hamiltonian of the form Eq. \ref{eqn:BdG-edge-Ham} with the spatial
coordinate in the interval $0<x<L$, must be supplemented by
antiperiodic boundary conditions on the Majorana field:
$\hat{\psi}(x=0) = - \hat{\psi} (x=L)$.  With this boundary condition,
the lowest energy edge mode will have energy $2\pi v/L$.  Of course,
in the thermodynamic limit this edge energy spacing vanishes, and one
has truly gapless edge excitations.

Next consider introducing a single $hc/2e$ vortex which is assumed to
be located at the center of the sample.  In the ``normal core" of the
vortex the order parameter vanishes, but the effects of the vortex are
``felt" by the Bogoliubov quasiparticles well outside of this region.
Indeed, upon exciting a fermionic Bogoliubov quasiparticle above the
bulk gap, and adiabatically transporting it around the vortex, the
fermionic quasiparticle will acquire a Berry's phase of $\pi$.  This
Berry's phase is equivalent to a sign change in the boundary condition
for the Bogoliubov quasiparticle upon encircling a vortex.  The edge
Majorana field encircling the outer edge of the sample, which in the
absence of the vortex has antiperiodic boundary conditions (due to the
$2\pi$ spinor rotation), will 
have periodic boundary conditions in the presence of a single bulk
vortex.  If $N_v$ denotes the number of bulk vortices, the boundary
condition on the edge Majorana fermion is
\begin{equation}
\hat{\psi}(x=0) = (-1)^{N_v + 1} \hat{\psi}(x=L)\  .
\end{equation}

For a single vortex, the periodic boundary condition on the Majorana
field implies that there will be one exact zero-energy eigenstate on
the edge, the zero-momentum state 
$ \hat{\psi}_{k=0} \equiv \hat{\psi}_{edge}$ independent of the
spatial coordinate.
Since all of the non-zero-energy states come in $\pm E$ pairs, it is
natural to anticipate the existence of a second zero-energy Majorana
mode associated with the vortex, and this is indeed the case.  To
illustrate this, consider modeling the vortex as a circular core
region of radius $r_{core}$, within which $V({\bf x}) - \mu < 0$, for
$|{\bf x}| < r_{core}$.  This depletes the fermion density within the
core region, in effect making a hole in the sampe, and creating an
internal edge running around the circumference of the hole.  Then,
just as for the outer sample edge, one expects an inner Majorana
chiral mode, described by a spinor with direction tangential to the
edge.  Moreover, the Berry's phase of $\pi$ experienced by the
Bogoliubov fermions upon encircling the vortex, will be ``felt" by the
chiral edge fermion encircling the core.  This leads to periodic
boundary conditions for this inner edge Majorana fermion, which will
have an exact zero-energy Majorana state, which we denote by
$\hat{\psi}_{vort}$.  The Majorana zero mode associated with the
vortex can be combined with the zero-energy Majorana mode on the
samples outer edge to define a zero-energy complex fermion,
\begin{equation}
\hat{a} \equiv \frac{1}{2} ( \hat{\psi}_{vort} + i \hat{\psi}_{edge} ),
\end{equation}
with $\hat{a}$ and $\hat{a}^\dagger$ satisfying canonical fermion
anticommutation relations, $\left\{ \hat{a}, \hat{a}^\dagger \right\}
= 1$.  Together, the vortex and edge zero modes thus constitute a
two-state system, corresponding to the two eigenvalues of
$\hat{a}^\dagger \hat{a} = 0,1$.  (For a finite system the parity of
the total electron number precludes one of the two states, implying a
unique ground state for the system with one vortex present.)  The
vortex ``quasiparticle" is thus entangled with the edge of the system,
despite the large spatial separation.

When multiple bulk vortices are present and spatially well separated
from one another, each will have an associated Majorana zero-energy
mode within its core.  For $N_v = 2N$ such vortices, the Majorana
zero modes can be combined to form $N$ complex fermions.  The
dimension of the zero-energy Hilbert space will thus be, $\Omega 2^{N-1}$, and for large $N_v$ scales as, $\Omega(N_v) \sim d^{N_v}$
with $d \equiv \sqrt{2}$.  Here $d$ is called the ``quantum" dimension
of the vortex quasiparticles. (We discuss this in greater depth in a
companion paper\cite{Fendley06c}.)  These vortex
quasiparticles have non-Abelian braiding statistics.

It is instructive to consider the process of bringing a vortex into
the system by passing it through the edge into the bulk of the sample.  Even
when the vortex is inside the sample and well away from the edge, it
leaves an imprint on the edge.  Specifically, this process dynamically
changes the boundary condition on the edge Majorana fermion field,
$\hat{\psi}(x)$ from periodic to antiperiodic (or vice versa) upon
encircling the sample.  But if the vortex is brought through the edge at
a particular spatial location, say $x$, the change in boundary
conditions must, in some sense, occur locally.  This can be made
precise by introducing an edge vortex field operator, denoted
$\hat{\sigma}(x)$, which satisfies
\begin{equation}
\hat{\sigma}(x) \hat{\psi}(x^\prime) \hat{\sigma}(x) = i \,\text{sgn}(x-x^\prime)
\hat{\psi}(x^\prime) .
\label{sigma_psi_com}
\end{equation}
This vortex quasiparticle field is Hermitian,
$\hat{\sigma}^\dagger = \hat{\sigma}$,
and squares to unity, $\hat{\sigma}(x) \hat{\sigma}(x) = 1$.
Thus Eq.~(\ref{sigma_psi_com}) can be re-expressed as, $\hat{\sigma}(x) \hat{\psi}(x^\prime)
= i \,\text{sgn}(x-x^\prime) \hat{\psi}(x^\prime)\hat{\sigma}(x)$.

We have employed the notation $\hat{\sigma}$ to denote this
boundary-condition-changing field, because of an intimate connection
between the edge theory of the $p_x+ip_y$ superconductor and a
one-dimensional quantum transverse Ising model tuned to criticality.
The $\sigma$ field is closely related to the Ising spin operator. We
will review the connection between Majorana fermions and the Ising
model in the next section.

\section{Majorana Fermions and the Ising Model}
\label{sec:Ising}

In the previous section, we saw how the edge modes for a
$p+ip$ superconductor are described by a free chiral Majorana
fermion. Near the end of the section, we saw that when a vortex passes
through an edge, it changes the Majorana fermion
boundary conditions from periodic to antiperiodic and vice-versa.
Such a process could be handled by introducing an edge vortex operator
which effects this change of boundary condition. As we
describe in this section, such an operator is closely related
to the spin field of the $2D$ critical Ising model.
We will review some key properties of chiral Majorana fermions
and their relation to the Ising model \cite{Lieb64}.
Excellent reviews of the two-dimensional
Ising lattice model and the Ising conformal field theory can be found
respectively in Ref.\ \onlinecite{mccoy73} and Ref.\
\onlinecite{ginsparg89}, so we will be brief.
In the next section, we will show how the
edge modes for the Moore-Read Pfaffian state are also described
by such a fermion (plus a free chiral boson), drawing on the notation
and terminology introduced here. 

The degrees of freedom in the Ising model are classical ``spins''
taking values $+$ or $-$ on the sites of some lattice.
However, as shown in Ref. \onlinecite{Lieb64}, the $2D$
Ising model can be reformulated as a theory of free fermions
on the lattice; they become massless at the critical point.
In the continuum limit, the model is described by a free
massless Majorana fermion.
The correlations of this fermion are, therefore, simple. However, the
map from the Ising spins to the fermionic variables is {\em non-local}.
The spin field introduces a branch cut for the fermions terminating
at the point at which it acts. Correlators of spins are therefore highly non-trivial.

To proceed further, it is useful to rotate Euclidean
time to real time, and obtain a Lorentz-invariant 1+1 dimensional
field theory. At the critical point, the modes of the fermion are either
right-moving or left-moving, which means the corresponding fields
${\psi}$ and $\overline{\psi}$  depend only on
$(vt-x)$ or $(vt+x)$ respectively. Since we are interested in
describing the edge modes for a disc, we take space to be periodic,
identifying $x=x+2\pi R$. Spacetime is thus a cylinder.

It is often more convenient to study conformal field theory on the
punctured plane instead of the cylinder. This can easily be done by
taking advantage of the conformal invariance of critical points and
performing a conformal transformation to the complex coordinates $z=e^{(vt-ix)/R}$
and $\overline{z}=e^{(vt+ix)/R}$. It is usually easiest to compute a
given correlator on the plane, and then do a conformal
transformation to the cylinder. One thing to note is that the
transformation between the cylinder and the plane changes the boundary
conditions on the fermion from periodic (antiperiodic) around the cylinder
to antiperiodic (periodic) around the puncture at the origin of the plane.
In complex coordinates, the action of the 1+1 dimensional critical
Ising field theory is then
\begin{equation}
{S}_\text{Ising} = -i\int dz\,d\overline{z}
[ {\psi}(x) \partial_{\overline{z}}  {\psi}(z) -
\overline{\psi}(\overline{z}) \partial_z  \overline{\psi}(\overline{z})  ] .
\label{SIsing}
\end{equation}
Going off the critical point
corresponds to adding a mass term $\propto \psi\overline{\psi}$ to
this field theory. Since the theory is quadratic in $\psi$ both on and
off the critial point, any correlators involving the fermions can
easily be computed.

The spin field $\sigma(z,\overline{z})$, on the other
hand, is a {\em twist field} for the fermions.
A twist field at a given spacetime location puts
a puncture there, so that the fermion boundary conditions around the
puncture are changed from periodic to antiperiodic. We thus demand
that the operator product of the twist field with the fermions be of
the form
\begin{eqnarray}
\psi(z)\sigma(w,\overline{w}) &\sim& \frac{1}{(z-w)^{1/2}}
\mu(w,\overline{w})\\
\overline{\psi}(\overline{z})\sigma(w,\overline{w})
&\sim& \frac{1}{(\overline{z}-\overline{w})^{1/2}} \mu(w,\overline{w})
\label{spsiOPE}
\end{eqnarray}
where $\mu$ is another field with the same dimension as $\sigma$. $\mu$
turns out to be the continuum limit of the Kramers-Wannier
dual of the spin field, which
is known as the {\em disorder field}. {}From (\ref{spsiOPE})
we see if we rotate $z$ around $w$ by an angle of $2\pi$, we pick
up a factor $(e^{\pm 2\pi i})^{-1/2}=-1$. In other words, the twist
field creates a square-root branch cut in fermion correlators. Thus the twist
field is non-local with respect to the fermions.

To change the boundary conditions for all the fermions,
one merely places twist fields $\sigma(0,0)$ and $\sigma(\infty,\infty)$ at the
origin and spacetime infinity of the punctured plane. (These points
correspond respectively to $t\to-\infty$ and $t\to+\infty$ on our
original spacetime cylinder; including these two points makes
spacetime topologically a sphere.) This creates a branch cut from the
origin to infinity on the plane, so that the fermions pick up a
minus-sign change in their boundary conditions.
We find, for instance, that for periodic (P) and anti-periodic (AP)
boundary conditions around the origin:
\begin{eqnarray}
\left\langle \psi(z)\psi(w)\right\rangle_\text{P} &=& \frac{1}{z-w}\cr
\left\langle \psi(z)\psi(w)\right\rangle_\text{AP} &=&
\left\langle\sigma(\infty,\infty) \psi(z)\psi(w) \sigma(0,0)\right\rangle_\text{P}\cr
&=& \frac{1}{z-w}\,\cdot\, \frac{1}{2}\left(\sqrt{\frac{z}{w}}+\sqrt{\frac{w}{z}}\right)
\end{eqnarray}
Using the latter correlation function, we can compute the operator
product expansion of $\sigma(z,\overline{z})$ with the energy-momentum
tensor, $T=\frac{1}{2}:\psi\partial\psi:$. From this operator product,
we can deduce that the right and left scaling dimensions of $\sigma(z,\overline{z})$
are $(\frac{1}{16},\frac{1}{16})$, for a total scaling dimension of $1/8$
(see, e.g. Ref \onlinecite{ginsparg89} for details).
By scaling, this gives Onsager's famous result that
$\eta=1/4$ in the $2D$ Ising model.

To obtain the correlation function of an arbitrary number of
spin fields, $\langle\sigma\sigma\ldots\sigma\rangle$,
we need to compute the ratio of the fermion partition function
in the presence of the corresponding branch cuts with the
partition function without any branch cuts. However, this
is a very difficult calculation in general. Instead, we can
use the powerful constraints which follow from two-dimensional
conformal invariance, which holds at the critical point.

In two dimensions, conformal transformations take
the form $z\rightarrow f(z)$,
$\overline{z}\rightarrow \overline{f}\left(\overline{z}\right)$,
where $f$ and $\overline{f}$ are arbitrary analytic functions.
Not only do these transformations decompose into independent
right and left transformations $f$ and $\overline{f}$, but
the algebra of infinitesimal transformations of this form
is infinite-dimensional -- two copies of the Virasoro
algebra, one for $z$ and one for $\overline{z}$ (see
refs. \onlinecite{Belavin84,ginsparg89} for details).
Consequently, operators and states can be organized in
representations of these two independent algebras.

The independence of these two algebras leads to
separate constraints for the $z$ and $\overline{z}$ dependence
of correlation functions. This naturally leads one to consider
the two chiralities separately. In general, there is no
local action for the chiral part of a conformal field theory
by itself, so the chiral theory must be considered purely algebraically.
However, in the case of the Ising model, there {\em is}
a local action for the right-moving part of the Ising model,
which only has $z$ dependence:
\begin{equation}
{S}_\text{chiral Ising} = -i\int dz\,d\overline{z}\:
{\psi}(x) \partial_{\overline{z}}  {\psi}(z)  
\label{SIsing-chiral}
\end{equation}
(and there is a similar action for the left-moving part alone).
Of course, in the context of edge excitations
of a $p+ip$ superconductor (as we saw in the last section)
or of a $\nu=\frac{5}{2}$ quantum Hall state (as we will
see in the next section), the chiral theory (\ref{SIsing-chiral})
itself actually interests us.
The fields in this theory can be organized in representations
of a single copy of the Virasoro algebra, corresponding
only to the transformations $z\rightarrow f(z)$ (since there
is no $\overline{z}$ dependence at all). The chiral spin field
$\sigma(z)$, which does not appear at all in the action
(\ref{SIsing-chiral}), is best understood in just such a way.

In a ``rational'' conformal field
theory, like the Ising conformal field theory, all states in the theory
can be found by acting with symmetry generators on a finite number of
states. The fields which create these special states are known
as ``primary'' fields. In other words, for each primary field,
there is a corresponding irreducible representation of the
Virasoro algebra (or a larger enveloping algebra),
whose states are obtained by acting
with all elements of the algebra on the state created by the primary field.
In the context of edge states, these primary fields
correspond to the different possible topological charges which
can be at the edge (they must, of course, be accompanied by
compensating topological charges in the bulk). By acting with
symmetry generators, we produce all possible generalized
oscillator excitations (such as edge magnetoplasmons)
`on top of' these topological charges.

For the chiral Ising model, there are just three primary fields,
which are the identity field $I$, the twist field $\sigma$, and the fermion
$\psi$. These three primary fields correspond to three irreducible
representations of the Virasoro algebra. Hence, the product
of any two such representations can be decomposed into
the sum of irreducibles. In the Ising model, the corresponding 
{\em fusion rules} are
\begin{eqnarray}
\label{eqn:Ising-fusion-rules}
\sigma \cdot \sigma &=& I + \psi\cr
\sigma \cdot \psi &=& \sigma\cr
\psi \cdot \psi &=& I
\end{eqnarray}
Of course, the product of any representation with the identity is the
representation itself. In terms of operators or fields, the fusion
rules amount to the statement that the primary fields on the
right-hand-side appear in the operator product expansion of the two
fields on the left.

These fusion rules for representations
correspond precisely to the rules for combining topological
charges in the bulk. Two nearby Majorana fermions in the bulk
of a $p+ip$ superconductor
are topologically equivalent to the ground state (i.e. the absence
of a topologically non-trivial excitation) as far as a distant quasiparticle
is concerned. On the other hand, two nearby vortices can either
be topologically equivalent to the ground state or to a single neutral fermion.
(These are the two states of the topological qubit which the two vortices
form.)

The chiral $\sigma(z)$ field, with scaling dimension $1/16$,
is largely the subject of this paper. In the Ising model context,
it would only be considered at an intermediate step of a calculation.
The non-chiral field $\sigma(z,\overline{z})$ is the field which
is really of interest in the Ising model. It is a primary field under
both the right- and left-handed Virasoro algebras, but it is {\it not}
simply the product of the right-handed chiral field $\sigma(z)$
with its left-handed partner. One can deduce this, for instance,
from the operator product expansion:
$$
\sigma(z,\overline{z})\sigma(w,\overline{w}) = \frac{1}{|z-w|^{1/4}} +
\frac{i}{2} |z-w|^{3/4} \psi(w)\overline{\psi}{\overline{w}}) +\dots.
$$
This expansion does not factor
into the product of right- and left-handed copies of
(\ref{eqn:Ising-fusion-rules}).  Therefore, the correlation functions
of the chiral field $\sigma(z)$ cannot be simply obtained by factoring
the $z$ and $\overline{z}$ dependence of the correlators of
$\sigma(z,\overline{z})$. Further subtleties must be dealt with, as we
discuss in section \ref{sec:conformal-blocks}.

\section{Edge excitations of the $\nu=5/2$ state} 
\label{sec:edgemodes} 
 
In this section, we will derive the form of
the theory of edge excitations of a $\nu=5/2$ droplet, 
assuming that the lowest Landau level
(of both spins) is filled and the first excited Landau level
is in the universality class of the
Moore-Read Pfaffian quantum Hall state. To do this,
we give the explicit form of wavefunctions for 
the edge excitations. Let us follow Milovanovic 
and Read \cite{Milovanovic96} and take the Hamiltonian 
to be the three-body interaction for which the Moore-Read
state is the exact ground state \cite{Greiter92} 
together with a confining potential which simply gives 
an energy proportional to the increase in angular momentum, 
$E \propto \Delta M$. Neither of these is realistic, but 
they make the counting of edge states easy, and the 
universal properties will not depend on these details. 
 
The Moore-Read wavefunction \cite{Moore91,Greiter92} for
filling fraction $\nu=1/m$ ($m$ even for fermions; odd for
bosons) is
\begin{equation} 
\Psi_0  ~=~ \prod_{j<k} (z_j - z_k)^m 
\prod_j e^{- |z_j|^2/4 } 
\cdot {\rm Pf}\!\left( \frac{1}{z_j - z_k }\right)~. 
\label{grdstate} 
\end{equation}
As opposed to the last section, $z$ here is a complex coordinate for
two-dimensional space, not 1+1-dimensional spacetime. 
The Pfaffian is the square root of the determinant of 
an antisymmetric matrix or, equivalently, the 
antisymmetrized product over pairs 
\begin{equation} 
\label{eqn:Pfaff-def} 
{\rm Pf}\!\left( \frac{1}{z_j - z_k }\right) = 
{\cal A}\left(\frac{1}{{z_1}-{z_2}}\frac{1}{{z_3}-{z_4}}\ldots\right) 
\end{equation} 
We will assume for now that there is an even number 
of electrons in the system and consider the odd 
electron number later. The form (\ref{eqn:Pfaff-def}) 
is strongly reminiscent of the real-space 
form of the BCS wavefunction. 
Indeed, the Moore-Read state may be viewed as a quantum Hall state 
of $p$-wave paired fermions \cite{Read00,Ivanov01,Stern04}. 
At $\nu=5/2$, we take $m=2$ for the electrons in the $N=1$ 
landau level. Other even-denominator quantum Hall 
states of electrons could be described by $m$ even. 
Quantum Hall states of bosons (e.g. cold atoms 
in rotating traps) would correspond to $m$ odd. 
 
There are $3m$ topologically-distinct quasiparticle types 
in this state which we will enumerate below. 
On a closed surface, the total topological charge must be trivial. 
In a system with boundaries, the total topological 
charge in the bulk is equal to the topological charge at 
the boundaries. Therefore, the Moore-Read 
state on a disk has $3m$ different 
sectors of edge excitations, corresponding to the 
different possible topological charges at the edge. 
 
There are sectors corresponding to different 
numbers of Laughlinesque charge $e/m$ quasiparticles 
in the bulk: 
$$
\Psi  = 
\prod_{i} {z_i}^n \, 
\prod_{j<k} (z_j - z_k)^m \prod_j e^{- |z_j|^2/4 }\,
{\rm Pf}\!\left(\frac{1}{z_j - z_k}\right). 
$$
These different charge sectors correspond to 
the different sectors (or primary fields) of a 
chiral boson $\phi \equiv \phi + 2\pi \sqrt{m}$: 
$e^{in\phi/\sqrt{m}}$, $n=0,1,\ldots,m-1$.
As in the case of the Laughlin states, in 
each of these sectors there are edge excitations 
which correspond to the 
multiplication of (\ref{grdstate}) by a symmetric polynomial 
$S\left({z_1},{z_2},\ldots,{z_N}\right)$: 
$$ 
\Psi  ~=~ S\left({z_1},{z_2},\ldots,{z_N}\right)\, 
\prod_{i} {z_i}^n \Psi_0
$$
The low-degree symmetric polynomials are in one-to-one 
correspondence with the oscillator modes of a free 
chiral boson \cite{Wen91}: 
\begin{equation} 
{\cal L}^{\rm charge} =\frac{1}{4 \pi} \partial_x 
\phi(\partial_t+{v_c}\partial_x)\phi \ .
\label{eqn:edge-boson} 
\end{equation}

These are the only edge excitations for the Laughlin states, but the
Moore-Read state has fermionic edge excitations as well.  Consider the
following states for $F$ even:
\begin{multline} 
\label{eqn:Majorana-wfcns} 
\Psi  ~=~ \prod_{i} {z_i}^r \, 
\prod_{j<k} (z_j - z_k)^m 
\prod_j e^{- |z_j|^2/4 }\: \times \\ 
\qquad{\cal A}\left({z_1^{p_1}}{z_2^{p_2}}\ldots {z_F^{p_F}} 
\frac{1}{z_{F+1}-z_{F+2}}\frac{1}{z_{F+3}-z_{F+4}}\ldots\right)~. 
\end{multline} 
The antisymmetrization requires that we take 
$0\leq{p_1}<{p_2}<\ldots<{p_F}$. Therefore, 
there is an exclusion principle for these excitations: 
we are populating fermionic edge modes with 
neutral fermions obtained by breaking pairs (they are 
neutral because the charge density 
is unchanged). 
The angular momentum increase is: 
\begin{equation} 
\label{eqn:anti-periodic} 
\Delta M = {\sum_i} \left({p_i}+\frac{1}{2}\right) 
\end{equation} 
These excitations are in one-to-one correspondence 
with the basis states of a Majorana fermion: 
$$\psi_{-{p_F}-\frac{1}{2}}\ldots 
\psi_{-{p_2}-\frac{1}{2}}\psi_{-{p_1}-\frac{1}{2}} 
|0\rangle 
$$
with Lagrangian: 
\begin{equation}
\label{eqn:L-neutral} 
{\cal L}^{\rm neutral} = 
i \psi(\partial_t+{v_n}\partial_x)\psi 
\end{equation} 
{}From (\ref{eqn:anti-periodic}), we see that 
$\psi$ has angular momentum quantized in half-integers 
in the sectors $e^{in\phi/\sqrt{m}}$, $n=0,1,\ldots,m-1$. 
 
Breaking pairs isn't the only way to populate these 
modes. We could also add an electron, 
so that the electron number is now odd. The ground 
state wavefunction of lowest angular momentum is 
\begin{multline} 
\Psi  ~=~ \prod_{i} {z_i}^r \, 
\prod_{j<k} (z_j - z_k)^m 
\prod_j e^{- |z_j|^2/4 }\: \times \\ 
{\cal A}\left({z_1^0} 
\frac{1}{{z_2}-{z_3}}\frac{1}{{z_4}-{z_5}}\ldots\right)~. 
\end{multline} 
We have now added a neutral 
fermion to the system, giving us the odd fermion 
number sectors $\psi\,e^{ir\phi/\sqrt{m}}$, 
with $r=0,1,\ldots,{m-1}$. 
We can, of course, multiply by symmetric 
polynomials to obtain bosonic oscillator excitations 
in these sectors as well. We can also break pairs as in 
(\ref{eqn:Majorana-wfcns}) -- but 
with $F$ now odd -- in order to populate an arbitrary 
odd number of fermionic modes. 
 
The paired nature of the Moore-Read state allows for 
quasiparticles carrying half of 
a flux quantum and, therefore, charge $1/2m$. 
A wavefunction for a two-quasihole state may be written 
by exploiting the Pfaffian factor in (\ref{grdstate}) 
to split a Laughlin quasihole into two half-flux-quantum 
quasiholes at $\eta_1$ and $\eta_2$: 
\begin{multline} 
\nonumber
\Psi  ~=~ \prod_{j<k} (z_j - z_k)^m
\prod_j e^{- |z_j|^2/4 }\times\\ 
{\rm Pf}\!\left(  \frac{\left({z_j}-{\eta_1}\right) 
\left({z_k}-{\eta_2}\right)+{z_j}\leftrightarrow{z_k}}{z_j - z_k}  
\right)~. 
\end{multline} 
If we take $\eta_1$ to infinity and $\eta_2$ to the origin, 
we have a wavefunction for a state with one half-flux 
quantum quasihole: 
$$
\Psi  ~=~ \prod_{j<k} (z_j - z_k)^m \prod_j e^{- |z_j|^2/4 } 
\: {\rm Pf}\!\left(\frac{{z_j}+{z_k}}{{z_j} -{z_k}}\right)~. 
$$
The extra factor of ${z_j}+{z_k}$ in the numerator 
gives each pair an additional 
unit of angular momentum. Majorana fermion edge excitations 
in this sector have wavefunction
\begin{multline} 
\nonumber
\label{eqn:Majorana-wfcns} 
\Psi  ~=~ \prod_{i} {z_i}^s \, 
\prod_{j<k} (z_j - z_k)^m 
\prod_j e^{- |z_j|^2/4 }\: \times \\ 
\qquad{\cal A}\left({z_1^{p_1}}{z_2^{p_2}}\ldots {z_F^{p_F}} 
\frac{z_{F+1}+z_{F+2}}{z_{F+1}-z_{F+2}} 
\frac{z_{F+1}+z_{F+2}}{z_{F+3}-z_{F+4}}\ldots\right)~. 
\end{multline} 
As a result of the extra angular momentum of 
each pair, the angular momenta of these excitations 
takes integral values: 
\begin{equation} 
\label{eqn:periodic} 
\Delta M = {\sum_i} {p_i} 
\end{equation} 
Therefore, a half-flux quantum quasiparticle 
has the effect of changing the quantization condition 
on fermion momenta from integer to half-integer 
values, in addition to the electrical charge it carries. 
Therefore, this is the $\sigma\, e^{i\phi/2\sqrt{m}}$ sector. 
The Ising spin field $\sigma$ introduces 
a branch cut for fermions $\psi$, thereby shifting their 
angular momenta by half a unit. We discuss correlation functions of
$\sigma$ in detail in section \ref{sec:conformal-blocks}.
 
Of course, we can also have an additional $s$ quasiholes 
at the origin, corresponding to topological charges 
$\sigma e^{(2s+1)i\phi/2\sqrt{m}}$: 
$$
\Psi  ~=~ 
\prod_{i} {z_i}^s \, 
\prod_{j<k} (z_j - z_k)^2 \prod_j e^{- |z_j|^2/4 } 
\: {\rm Pf}\!\left(\frac{{z_j}+{z_k}}{{z_j} -{z_k}}\right)~. 
$$
 
These $3m$ sectors can essentially be divided 
into $m$ different charge sectors and $3$ neutral 
sectors, where the non-Abelian structure lies. The one 
subtlety is that the space of states is not simply a tensor 
product of charged and neutral sectors but only includes those
invariant under the combined transformation
$\sigma\rightarrow -\sigma$, 
$\phi\rightarrow \phi + 2\pi\sqrt{m}$. 

To summarize, the edge excitations of a droplet of
$\nu=1/m$ Moore-Read liquid obey the Lagrangian
\begin{multline}
{{\cal L}_\text{edge}}(\psi,\phi) = {{\cal L}_\text{fermion}}(\psi)
+ {{\cal L}_\text{boson}}(\phi)\\
= i \psi({\partial_t}+{v_n}\partial_x)\psi
+ \frac{1}{2 \pi} {\partial_x}\phi({\partial_t}+{v_c}{\partial_x}\phi) 
\label{Ledge}
\end{multline}
with $\phi\equiv \phi + 2\pi\sqrt{m}$.
The neutral and charge velocities ${v_n}$, ${v_c}$ are, in general,
different, and one expects that ${v_n}<{v_c}$. The normalization above
is such that the operator $e^{ia\phi}$ has scaling
dimension ${a^2}/2$, or equivalently that the two point function,
$\langle e^{ia\phi(\tau)} e^{-ia\phi(\tau^\prime)} \rangle \sim |\tau - \tau^\prime|^{-2 d_a}$, evaluated for the Gaussian Lagrangian ${\cal L}_{boson}(\phi)$ decays as a power law with $d_a = a^2/2$.
The different primary fields,
i.e. topologically distinct quasiparticles, are:
\begin{eqnarray}
\Phi_{q/m} &=& e^{iq\phi/\sqrt{m}}\cr
\Phi_{\psi,q/m} &=& \psi\, e^{iq\phi/\sqrt{m}}\cr
\Phi_{(2r+1)/2m} &=& \sigma\, e^{i(2r+1)\phi/2\sqrt{m}}
\end{eqnarray}
with $q,r=0,1,\ldots,m-1$. There are also quasiparticles
with $q\geq m$ or $r\geq m$, but these do not correspond to primary fields
(they are, instead, descendant fields) because a quasiparticle with
$q\geq m$ has the same topological properties as
the quasiparticle with $q\rightarrow q(\text{mod }m)$
above, and similarly for $r\geq m$.
$\Phi_{0} = 1$ is the identity operator, which is topologically
trivial and has the same quantum numbers as the vacuum.
Other topologically trivial operators are descendants of the identity.
There is one such descendant of the identity
which is particularly important physically, namely the electron
(fermionic for $m$ even):
\begin{equation}
\Phi_\text{el} = \psi \,e^{i\phi\sqrt{m}}
\end{equation}
Two other topologically trivial operators which will
interest us later are an operator which
annihilates a charge-$2$ boson, which we will
interpret as a Cooper pair,
\begin{equation}
\Phi_\text{pair} = e^{2i\phi\sqrt{m}}
\end{equation}
and the fermion kinetic energy operator
\begin{equation}
\Phi_{\psi,\text{kin}} = \psi{\partial_x}\psi
\end{equation}
which we can interpret as the creation/annihilation
operator for a $p$-wave pair of neutral fermions.
For the $\sigma_{xy}=\left(2+\frac{1}{2}\right)\,\frac{e^2}{h}$
quantum Hall state, we take $m=2$ in
the above formulas.

\section{The Point Contact} 
\label{sec:point-contact}

\begin{figure}[t!]
\centerline{\includegraphics[width=2.85in]{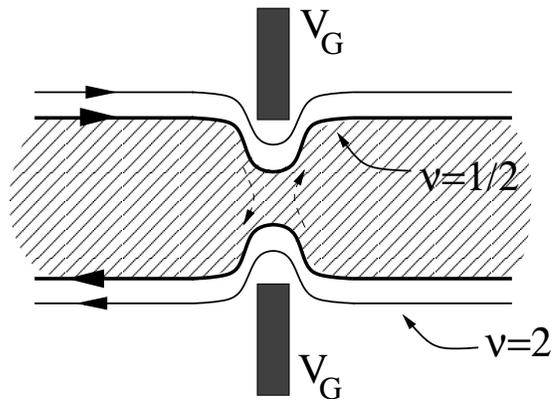}}
\caption{A voltage $V_G$ applied to gates on either side of a Hall bar
forms a constriction, causing tunneling between the edges.  For a weak
constriction in a $\nu=5/2$ state, quasiparticles can tunnel between
the edges of the half-filled first excited Landau level. Tunneling
between the integer quantum Hall edges of the filled lowest Landau
level can be neglected because these edges are further apart.}
\label{fig:Hall-bar}
\end{figure}

\begin{figure}[t!]
\centerline{\includegraphics[width=2.80in]{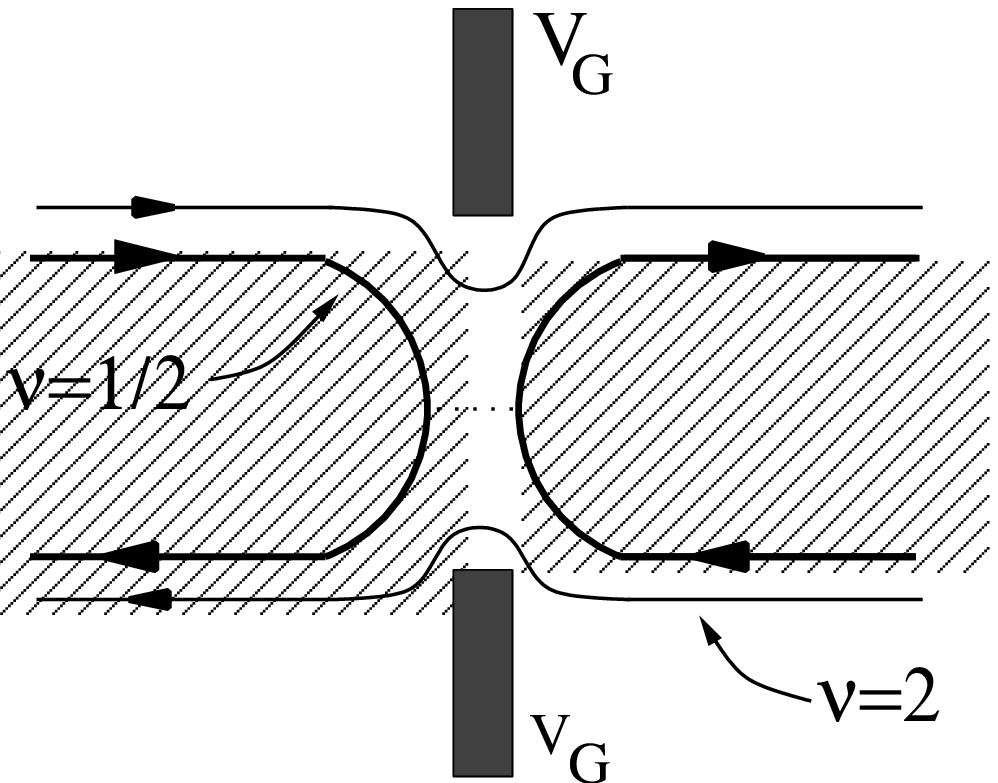}}
\caption{When a large gate voltage $V_G$ is applied, the
quantum Hall droplet is broken in two. Electrons can tunnel
between the two droplets. At $\nu=5/2$, it is the $\nu=1/2$ droplet
in the first excited Landau level which is broken in two. The
$\nu=2$ integer quantum Hall droplet remains unbroken.}
\label{fig:broken-bar}
\end{figure}

A voltage $V_G$ applied to the gates on either side of a Hall droplet
effectively pinches the droplet, as illustrated in figure
\ref{fig:Hall-bar}. Quasiparticle tunneling between the edges, which
is negligible when they are far apart, will now become important in
the vicinity of the constriction. If the gate voltages are large, then
the Hall bar will be cut in two by the gates so that there are two
Hall droplets, as depicted in figure \ref{fig:broken-bar}.  On the
other hand, if the droplet is not pinched too strongly, we might
naively expect that the rate at which quasiparticles tunnel between
the top and bottom edges will be small.  However, as we will discuss
in detail, a weak pinch will always become effectively stronger as the
temperature is decreased, until it reaches the limit of a Hall droplet
which is broken in two at zero temperature.

In the strong constriction limit, there is vacuum separating the
two droplets, so only electrons and excitations which are made up of
several electrons -- i.e.\ topologically trivial excitations -- can
tunnel between the left and right droplets. (In a 
$\nu=2+\frac{1}{2}$ Moore-Read Pfaffian state,
we want to consider the case in which there is $\nu=2$
integer quantum Hall liquid, which also does not support
fractional excitations, between the two $\nu=5/2$
droplets.) Therefore, for example,
a charge-$1$ boson cannot tunnel between the two droplets, for
$m$ even. Even though its charge is integral, it braids non-trivially with
a charge $1/2m$ excitation.

Let us first consider the action describing a strong constriction.
There are two edges, a counterclockwise edge on the left,
which we will denote by $L$; and a clockwise edge on the right,
which we will denote by $R$. We mark distance along both
of these edges with a spatial coordinate $x$.
The point contact is at $x=0$. The most important operators
coupling the two edges tunnel an electron, a pair of electrons, or
a pair of neutral fermions from the point $x=0$ on one edge
to $x=0$ on the other:
\begin{multline}
\label{eqn:strong-coupling-general}
S_\text{strong} = \int d\tau\, \left[\int dx\,\left({\cal L}_\text{edge}({\psi_L},{\phi_L})
+ {\cal L}_\text{edge}({\psi_R},{\phi_R})\right)\right.\\
+ t_\text{el}\,
\Phi^\dagger_{\text{el},R}(0)\Phi_{\text{el},L}(0) 
+ t_\text{pair}\,
\Phi^\dagger_{\text{pair},R}(0)\Phi_{\text{pair},L}(0) \\
+ t_{\psi,\text{kin}}\,
\Phi^\dagger_{\psi,\text{kin},R}(0)\Phi_{\psi,\text{kin},L}(0)
+ \text{h.c.}\Big]
\end{multline}
A term coupling the kinetic energies of the charged bosons could also
appear in this action, but it is less important than $
t_{\psi,\text{kin}}$ at low temperatures and is not particularly
important in the analysis which follows. 

As for tunneling between Abelian quantum Hall edges, one can readily
define a renormalization group (RG) transformation which leaves the
edge Lagrangians invariant\cite{Kane92}.  The edge Lagrangan is an RG
``fixed point". To read off the scaling dimensions of the operators,
it is convenient to write the action
(\ref{eqn:strong-coupling-general}) in terms of the boson introduced
in section \ref{sec:edgemodes}:
\begin{multline}
S_{\hbox{strong}} =  \int d\tau\,\left[ \int dx\,\left({\cal L}_\text{edge}({\psi_L},{\phi_L})
+ {\cal L}_\text{edge}({\psi_R},{\phi_R})\right)\right.\\
+t_\text{el}\, i{\psi_R}(0){\psi_L}(0)
\cos(({\phi_R}(0)-{\phi_L}(0))\sqrt{m})\\
+t_\text{pair}  \,\cos(2({\phi_R}(0)-{\phi_L}(0))\sqrt{m})\\
+t_{\psi,\text{kin}}\, \left({\psi_R}(0){\partial_x}{\psi_R}(0)\right) \,
\left({\psi_L}(0){\partial_x}{\psi_L}(0)\right)\Big]
\end{multline}
Under this RG transformation, the leading terms for the ``flows" of
the tunneling operators are:
\begin{eqnarray}
\label{eqn:strong-coupling-RG-general}
\frac{d\,}{d\ell}t_\text{el} &=& -m \,t_\text{el}\cr
\frac{d\,}{d\ell}t_\text{pair} &=&
(1-4m)\,t_\text{pair}\cr
\frac{d\,}{d\ell}t_{\psi,\text{kin}} &=& 
-3\,t_{\psi,\text{kin}}
\end{eqnarray}
Since these operators are all irrelevant at the fixed point,
the limit of two decoupled droplets is stable to weak inter-droplet
tunneling.

The case $m=2$ is relevant to the half-filled
first excited Landau level in a Moore-Read
Pfaffian state at $\nu=2+\frac{1}{2}$. Here $t_\text{el}$ has RG eigenvalue
$-2$ and is the least irrelevant coupling
for both charge and energy transport.
Consequently, the $4$-terminal longitudinal resistance (see
appendix \ref{sec:four-terminal}) scales
as:
\begin{equation}
\label{eqn:low-T-generic}
R_{xx} - \frac{1}{10}\,\frac{h}{e^2} \sim - {t_\text{el}^2}\,{T^4}
\end{equation}
$t_{\psi,\text{kin}}$ is the first sub-leading irrelevant operator
coupling the two droplets, but it does not contribute to charge
transport.  The leading sub-dominant contribution to the charge
transport between the two droplets is thus from the pair tunneling
term, $t_\text{pair}$, which is strongly irrelevant.  If $t_\text{el}$
is tuned to zero, then
\begin{equation}
\label{eqn:low-T-non-generic}
R_{xx} - \frac{1}{10}\,\frac{h}{e^2} \sim -{t_\text{pair}^2}\,T^{14}
\end{equation}
In the case of a $p+ip$ superconductor, there is no charged mode,
so $t_{\psi,\text{kin}}$ is the only one of these three couplings which
can occur. It is the most relevant coupling between the edge
modes of two such superconductors.

Now we turn our attention to the case of a weak constriction.
In this case, quasiparticles can tunnel across the bulk of
a Hall droplet, as in figure \ref{fig:Hall-bar}.
In the $\nu=2+\frac{1}{2}$ case, we assume that
tunneling only occurs between the $\nu=\frac{1}{2}$ edges
and ignore tunneling between the $\nu=2$ integer quantum
Hall edges, which are further apart.
It is convenient to treat the top and bottom
edges of the bar as independent, but we must keep in mind that,
ultimately, they are the two edges of the same bar.
We will use the subscripts $a$ and $b$ for the top
and bottom edges so that ${\psi_a},{\phi_a}$ are the
fermion and boson operators
at edge $a$ and ${\psi_b},{\phi_b}$ are the corresponding operators
at edge $b$. As drawn in the figure, the $a$ modes are right-moving
and the $b$ modes are left-moving.

At the point contact, which we will assume is at $x=0$,
the two edges are coupled by quasiparticle tunneling.
There are no restrictions on what kind of quasiparticles
can tunnel at the point contact. In general, we must
consider not only primary fields but also all of their
descendants. However, descendant fields have higher
scaling dimensions than primaries; as a consequence,
the tunneling of descendants is strongly irrelevant. For instance,
the tunneling of electrons, electron pairs, and neutral fermion
pairs -- all due to descendant fields --
can occur not only between droplets but also across a droplet.
As we saw above, they are all irrelevant.

Hence, if we retain only the tunneling of primary fields,
the action will take the form
\begin{multline}
\label{eqn:tunneling-action}
S_{\hbox{weak}} = \int d\tau\, \left[\int dx\, \left(
{\cal L}_\text{edge}({\psi_a},{\phi_a})
+ {\cal L}_\text{edge}({\psi_b},{\phi_b})\right)\right.\\
+
{\sum_{r=0}^{m-1}}
\left( \lambda_{(2r+1)/2m} \Phi^{a, \dagger}_{(2r+1)/2m}(0) \Phi^b_{(2r+1)/2m}(0)
+ \text{h.c.}\right)\\
+
{\sum_{q=0}^{m-1}}
\left( \lambda_{\psi,q/m}\, i\Phi^{a,\dagger}_{\psi,q/m}(0)
\Phi^b_{\psi,q/m}(0) + \text{h.c.}\right)\\
\left.{\sum_{q=1}^{m-1}}
\left( \lambda_{q/m} \Phi^{a\dagger}_{q/m}(0) \Phi^b_{q/m}(0)
+ \text{h.c.}\right)\right]
\end{multline}
($q=0$ is omitted from the last sum because
it is simply the identity operator.)
The leading terms in the
RG equations for the couplings above are:
\begin{eqnarray}
\frac{d\,}{d\ell}\lambda_{(2r+1)/2m} &=& 
\left(\frac{7}{8}-\frac{(2r+1)^2}{4m}\right)\lambda_{(2r+1)/2m}\cr
\frac{d\,}{d\ell}\lambda_{\psi,q/m} &=&
-\frac{q^2}{m}\,\lambda_{\psi,q/m}\cr
\frac{d\,}{d\ell}\lambda_{q/m} &=& \left(1-\frac{q^2}{m}\right)\lambda_{q/m}
\end{eqnarray}
In this equation, we have used the known scaling dimension
of the chiral part of the Ising spin field, $1/16$.

We now specialize to the two cases of greatest experimental interest,
a possible Moore-Read Pfaffian state at $\nu=2+\frac{1}{2}$
and a $p+ip$ superconductor.
A Moore-Read Pfaffian state at $\nu=2+\frac{1}{2}$
corresponds to $m=2$. Keeping only the terms in the action
which are not irrelevant, we have:
\begin{multline}
\label{eqn:five-halves-action}
S = \int d\tau\,\left[\int dx\, \left(
{\cal L}_\text{edge}({\psi_a},{\phi_a})
+ {\cal L}_\text{edge}({\psi_b},{\phi_b})\right)\right.\\
+ \lambda_{1/2} \, \cos(({\phi_a}(0)-{\phi_b}(0))/\sqrt{2})
+ \lambda_{\psi,0}\, i{\psi_a}{\psi_b}\\
+ \lambda_{1/4} \,{\sigma_a} {\sigma_b}\,
\cos(({\phi_a}(0)-{\phi_b}(0))/2\sqrt{2})\Big]
\end{multline}
In this action we have labeled the $\sigma$ field in the same fashion
as the other fields: by an index $a$ and $b$ indicating which edge it
is on. However, these sigma fields can be entangled with each other as
well as with other fields, and so defining the action precisely
requires more information than just the $a,b$ labels. We discuss this
issue in depth in section \ref{sec:conformal-blocks}. The leading
weak-tunneling corrections derived in this section
are not affected by this (important) subtlety.
 
The RG equations to lowest order for the three couplings
in (\ref{eqn:five-halves-action}) are:
\begin{eqnarray}
\label{eqn:Pfaffian-RG}
\frac{d\,}{d\ell}\lambda_{1/2} &=& \frac{1}{2}\,\lambda_{1/2}\cr
\frac{d\,}{d\ell}\lambda_{\psi,0} &=& 0\cr
\frac{d\,}{d\ell}\lambda_{1/4} &=& \frac{3}{4}\,\lambda_{1/4}
\end{eqnarray}
Since $\lambda_{1/4}$ and $\lambda_{1/2}$ are relevant
couplings, the weak tunneling limit is unstable.
The longitudinal resistivity increases with decreasing temperature:
\begin{equation}
R_{xx}\sim  \lambda^2_{1/4}\,T^{-3/2}
\end{equation}
If $\lambda_{1/4}$ were tuned to zero, we would instead have
$R_{xx}~\sim ~\lambda^2_{1/2}/T$. Since both $\lambda_{1/4}$ and
$\lambda_{1/2}$ grow as the temperature is decreased, we expect
that the constriction will become strong and the droplet will be be
effectively broken into two droplets. In the subsequent sections we
will show that this is the case \cite{Fendley06a} and describe the
crossover between these two limits.

\begin{figure}[t!]
\centerline{\includegraphics[width=3.45in]{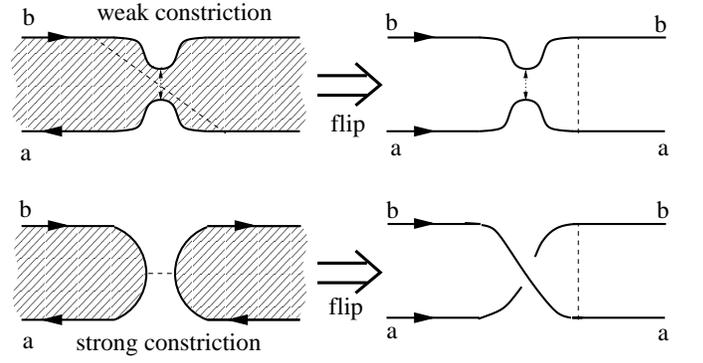}}
\caption{We redraw the point contact with the
bottom edge flipped so that the description is completely
chiral. In the strong constriction limit, an incoming $a$
mode becomes an outgoing $b$ mode and vice versa.
The dotted line with arrows on both ends
in the top two pictures represents the tunneling path
between the top and bottom edges in the weak constriction
limit. The dashed line in the bottom two pictures represents a tunneling
path from the left droplet to the right one. This dashed line is transposed
to the top two pictures to illustrate how these tunneling paths cross
in real space but are parallel in the flipped representation.}
\label{fig:flipped}
\end{figure}

In the preceding discussion, we have represented edge
excitations at the top and bottom edges of the droplet
as opposite chirality $(1+1)$-D theories. However, we
are not required to do so since tunneling occurs at only a
single point. We can exploit a trick used in the analysis of the
related (but distinct) problem of the Ising model with a defect line \cite{oshikawa97,leclair99}. If we flip the bottom edge, as shown in figure
\ref{fig:flipped}, then both edge modes are right-moving.
We now have the problem of a point defect in a purely chiral
theory. In the strong constriction limit, the incoming chiral modes
are exchanged as they pass through the contact at $x=0$,
as depicted in figure \ref{fig:flipped}.The (irrelevant) tunneling processes
which transfer electrons between the two droplets can
be implemented with chiral operators acting `downstream' from
the point contact. This is depicted by the dashed lines in
figure \ref{fig:flipped}. Note that when such an operator is transposed
back to the unflipped picture for the weak constriction limit
(the upper left of figure \ref{fig:flipped}), the dashed electron
tunneling path crosses the dotted quasiparticle tunneling path.
Since quasiparticle and electron tunneling operators commute,
the corresponding chiral operators in the upper right
of figure \ref{fig:flipped} also commute.

\begin{figure}[t!]
\centerline{\includegraphics[width=3.45in]{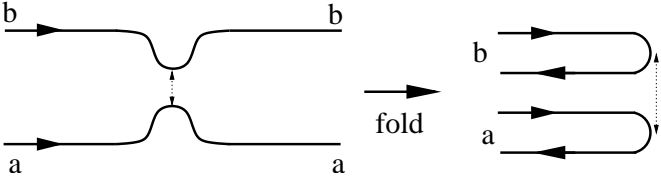}}
\caption{We can fold the $x>0$ half-plane
onto the $x<0$ half-plane so that
right-moving modes in the $x>0$ half-plane 
now become left-moving modes in the $x<0$ half-plane.
The resulting non-chiral modes are coupled only at the origin.}
\label{fig:folded}
\end{figure}

This `flipped' representation will prove to be more convenient,
as we will see in section \ref{sec:bosonization}. Even before
we get the real payoff in that section, however,
we can benefit from a minor simplification.
$\phi_a$ and $\phi_b$ are now both right-moving chiral bosons.
We can form their sum and difference:
${\phi_c}=\left({\phi_a}+{\phi_b}\right)/\sqrt{2}$,
${\phi_\rho}=\left({\phi_a}-{\phi_b}\right)/\sqrt{2}$. Only
${\phi_\rho}$ is affected by the point contact. ${\phi_c}$ decouples
completely, so we drop it from the action:
\begin{multline}
\label{eqn:flipped-five-halves-action}
S = \int d\tau\, dx\, \left( {{\cal L}_\text{fermion}}({\psi_a})
+ {{\cal L}_\text{fermion}}({\psi_b}) +
{{\cal L}_\text{boson}}({\phi_\rho})
\right)\\
+ \int d\tau\, \lambda_{1/2} \, \cos{\phi_\rho} 
+ \int d\tau\,\lambda_{\psi,0}\, i{\psi_a}{\psi_b}\\
+ \int d\tau\, \lambda_{1/4} \,{\sigma_a} {\sigma_b}\,
\cos({\phi_\rho}/2)
\end{multline}
We can now recast this problem as a boundary conformal
field theory problem by folding the $x>0$ half-plane
onto the $x<0$ half-plane, as shown in figure \ref{fig:folded}.
The $x>0$ part of right-moving modes
now become left-moving modes in the $x<0$ half-plane:
$\phi_{\rho R}(x)\equiv{\phi_\rho}(x)$, $\phi_{\rho L}(x)\equiv{\phi_\rho}(-x)$,
for $x<0$ and similarly for $\psi_{a,b}$.

In the case of a $p+ip$ superconductor, there is no charged
mode. Dropping irrelevant terms, the action in the `flipped'
representation is simply
\begin{multline}
\label{eqn:p+ip-action}
S = \int d\tau\,dx\, \left({{\cal L}_\text{fermion}}({\psi_a})
+ {{\cal L}_\text{fermion}}({\psi_b})\right) \\
+ \int d\tau\, \lambda_\psi\, i{\psi_a}{\psi_b}
+ \int d\tau\, \lambda_\sigma {\sigma_a}{\sigma_b}
\end{multline}
with
\begin{eqnarray}
\label{eqn:p+ip-RG}
\frac{d\,}{d\ell}\lambda_\psi &=& 0\cr
\frac{d\,}{d\ell}\lambda_\sigma &=& \frac{7}{8}\,\lambda_\sigma
\end{eqnarray}

\section{Conformal Blocks and Tunneling Operators} 
\label{sec:conformal-blocks}

In order to follow the crossover from the unstable weak tunneling
limit to the limit of two droplets, we need to go beyond lowest-order
perturbation theory in the relevant couplings described in the
previous section. In so doing, we see that the preceding discussion
requires refinement in one important respect. Terms of the form
${\sigma_a}{\sigma_b}$ in the above actions
(\ref{eqn:tunneling-action}) and (\ref{eqn:p+ip-action}) are not well
defined without additional information.  The technical reason is that
the correlation functions of the Ising spin field (as defined by
taking the continuum limit of the lattice model at its critical point,
or by using conformal field theory) do not factor into a product of a
right-moving and a left-moving part. 
For instance, the four-point function of the
non-chiral spin field $\sigma(z,\overline{z})$ on the plane is of the
form
\begin{equation} 
\langle \sigma(z_1,\overline{z}_1) 
\sigma(z_2,\overline{z}_2)\sigma(z_3,\overline{z}_3) 
\sigma(z_4,\overline{z}_4) \rangle = |{\cal F}_I(z)|^2 + |{\cal 
F}_\psi(z)|^2 
\label{fourpoint} 
\end{equation} 
where the {\em conformal blocks} ${\cal F}_I$ and ${\cal F}_\psi$
depend only on the $z$ coordinates with no $\overline{z}$ 
dependence. The sum in (\ref{fourpoint}) means
however that one {\em cannot} simply decompose 
$\sigma(z,\overline{z})$ into a product $ \sigma(z)\sigma(\overline{z})$. 
Therefore, the correlation functions
of, say, $\sigma_a$ are not well defined until more information is
provided. In the four-point case, the correlation function could be
${\cal F}_I$, ${\cal F}_\psi$, or even any linear combination of the
two.

The physical reason underlying this ambiguity is 
the reason why the problem is so interesting: the non-abelian 
braiding. Quasiparticles on the edge can be entangled with others, 
even if they are far away. Moreover, they also have 
non-trivial braiding with bulk quasiparticles: when a fermion goes 
around the disk, the sign the wavefunction picks up depends on the 
number of $\sigma$ quasiparticles in the bulk. (Thus the boundary 
conditions on the edge theory depend on whether the number of bulk 
quasiparticles is even or odd.)

These ambiguities are resolved by specifying the {\em fusion
channels} of the fields in addition to their positions in
spacetime. As we saw in (\ref{eqn:Ising-fusion-rules}, two chiral spin
fields can fuse either to the identity field $I$ or the fermion
$\psi$. What this means is that two spin fields form a two-state
quantum system, with $I$ and $\psi$ as the basis
elements. Consquently, conformal
blocks of four chiral spin fields form a two-dimensional vector space.
The conformal blocks in (\ref{fourpoint}) ${\cal F}_I$ and ${\cal
F}_\psi$ are those which the chiral fields at $z_1$ and $z_2$ fuse to
$I$ and $\psi$ respectively.

In this section, we define unambiguously the tunneling operators ${\cal
T}_\sigma$ and ${\cal T}_\psi$, which correspond respectively to
tunneling a $\sigma$ and $\psi$ quasiparticle across a $p+ip$
superconductor. This amounts to defining their conformal blocks
uniquely. In the subsequent sections, we exploit the work of ref.
\onlinecite{Moore88} to express these blocks in terms of correlators
of bosonic fields. This will enable us to derive the
surprising result that ${\cal T}_\sigma$ for a simple point contact is
equivalent to the interaction in the single-channel anisotropic Kondo
problem.

The effects of the non-Abelian statistics are independent
of the charged mode, so in this section, for simplicity, we study only
the $p+ip$ superconductor, where there are 
$\sigma$ and $\psi$ quasiparticles.  In the edge conformal field 
theory, all descendant fields are irrelevant, so the only fields we 
need to worry about are the $\sigma$ and $\psi$ primary fields. Since 
there is only one fusion channel for the edge fermion $\psi$, its 
tunneling operator is easy to define. 
Since the edge theory is conformally invariant, the correlations for any shape 
droplet can be obtained from those on a circular disk. We can then
label the position on the edge by an angular coordinate  $0\le
\theta<2\pi$ going around the circle.
For simplicity, we put the contact in the middle of the circular disc, 
so that it tunnels particles between $\theta=\pi/2$ and 
$\theta=3\pi/2$. Tunneling a fermion at Euclidean time $\tau$ then 
corresponds to the operator 
\begin{equation} 
{\cal T}_\psi(\tau)=\psi(\tau+i\pi/2)\psi(\tau+i3\pi/2) 
\label{psialpha} 
\end{equation} 
The fermion is free, so correlators of ${\cal T}_\psi$ are trivial to
compute.  Since $\psi$ is of dimension $1/2$, ${\cal T}_\psi$ is of
dimension 1, and is marginal.  The operator ${\cal T}_\sigma$ is
relevant, but we will show how the presence of ${\cal T}_\psi$ still
plays a crucial role in understanding the ``infrared fixed point",
which describes the low-temperature limit of two decoupled droplets.
(We refer to the opposite limit of a very weak pinch across the
Hall bar as the ``ultraviolet fixed point", since it describes the
behavior at temperatures much higher than the scale where the system
crosses over to two decoupled drops.)

The interesting complications occur for tunneling $\sigma$
quasiparticles. 
In order to properly define the tunneling operator
${\cal T}_\sigma$, we
need to account for the fact that our two edges
are, in fact, different sections of the same edge,
bounding a single Hall droplet. In figures
\ref{fig:Hall-bar}-\ref{fig:folded}, the edges going off to the left
are connected to each other, and likewise for those on the right.

The fact that the both point contacts are a single edge means that
the fields denoted in the last section by $\sigma_a$ and $\sigma_b$
are part of the same edge conformal field theory. When defining their
conformal blocks, we must therefore specify the appropriate fusion channels.
As shown in the fusion rules (\ref{eqn:Ising-fusion-rules}), two 
$\sigma$ fields can fuse to either the identity or the fermion. Tunneling 
a quasiparticle at Euclidean time $\tau$ involves two quasiparticle 
fields: one to annihilate a quasiparticle, and the other to create one 
on the other side.  To define this operator, we must therefore specify 
which fusion channel these two fields are in. {\it For a simple point 
contact, this must be the identity channel.} The reason is that simply
transferring a quasiparticle from one point to another nearby
point cannot create a neutral fermion. One 
can of course imagine more complicated physical situations, where it 
is possible for the tunneling to change fusion channels. For example, 
if there were an antidot with the point contact, the tunneling could 
be in the $\psi$ channel, with a compensating $\psi$ particle
created at the anti-dot. Consequently, a tunneling event in the $\psi$ channel 
corresponds to changing the topological charge
on the antidot by fusing with 
$\psi$ -- if the antidot were originally in the $I$ channel, the 
tunelling event would leave it in the $\psi$ channel, while if it were 
originally in the $\psi$ channel it would be left in the $I$ channel. 
Such point contacts can be analyzed using the formalism we develop 
here, but the physics becomes considerably more involved. We therefore 
focus in this paper on a simple point contact.

To explain in more detail what it means for the tunneling operator to 
be in the identity channel, we discuss its conformal blocks. A 
convenient pictorial representation of a chiral $\sigma$ field at 
$z_1$ is given by the trivalent vertices 
\begin{center} 
\begin{picture}(200,30) 
\put(20,3){$c$} 
\put(20,0){\line(1,0){40}} 
\put(40,0){\line(0,1){15}} 
\put(40,20){$1$} 
\put(90,5){or} 
\put(120,0){\line(1,0){40}} 
\put(140,20){$1$} 
\put(140,0){\line(0,1){15}} 
\put(157,3){$c$} 
\end{picture} 
\end{center} 
where $c$=$I$ or $\psi$. 
(In this figure, the $1$ refers to the spacetime point $z_1$
and must not be confused with the identity label, $I$.)
Such operators are generally known in the 
mathematical literature as ``chiral vertex operators''.  
In general, one labels all the legs of the vertex; here we adopt the 
convention that unlabeled 
legs correspond to the $\sigma$ channel.

A non-vanishing conformal block of $2n$ $\sigma$ fields located 
at $z_1\dots z_{2n}$ 
is then pictorially represented as 
\begin{center} 
\begin{picture}(250,30) 
\put(10,3){$I$} 
\put(30,20){$1$} 
\put(50,20){$2$} 
\put(90,20){$3$} 
\put(110,20){$4$} 
\put(67,3){$c_1$} 
\put(10,0){\line(1,0){125}} 
\put(30,0){\line(0,1){15}} 
\put(50,0){\line(0,1){15}} 
\put(90,0){\line(0,1){15}} 
\put(110,0){\line(0,1){15}} 
\put(125,3){$c_2$} 
\put(145,3){$\dots$} 
\put(165,3){$c_{n-1}$} 
\put(165,0){\line(1,0){65}} 
\put(190,0){\line(0,1){15}} 
\put(210,0){\line(0,1){15}} 
\put(178,20){$2n$--1} 
\put(210,20){$2n$} 
\put(227,3){$I$} 
\end{picture} 
\end{center} 
where $c_j$ represents the fusion channel for the first $2j$ particles; 
we must have $c_0=c_{2n}=I$ for the conformal block to be non-vanishing.  
This means that for a conformal block to be non-vanishing, the 
fusion channels for all the operators combined must be the 
identity. Each choice of $c_j=I,\psi$, with $j=1,\ldots,n-1$,
corresponds to a basis element for the 
vector space of $2n$-point conformal blocks. In ref. \onlinecite{Fendley06a},
we introduced the notation $[{m_1},\ldots,{m_n}]$ with
${m_i}=0,1$ for such a conformal block. The relation with the pictures above
is $c_j=I$ if $\sum_{i=1}^{j}{m_i} \equiv 0 (\text{mod} 2)$ and
$c_j=\psi$ if $\sum_{i=1}^{j}{m_i} \equiv 1 (\text{mod} 2)$. 
 
An arbitrary conformal block is of $2n$ $\sigma$ fields is therefore a
linear combination of these $2^{n-1}$ basis elements. Again, we see
that the quantum dimension of $\sigma$ is $\sqrt{2}$.  Conformal
blocks with four $\sigma$ quasiparticles form a two-dimensional vector
space. Thus they effectively form a two-state quantum system, which
can be used as a qubit in a topological quantum computer.

In this pictorial notation, we then have for the tunneling operator of 
a simple point contact 
\begin{equation} 
\begin{picture}(110,30) 
\put(0,3){${\cal T}_\sigma\equiv$} 
\put(30,3){$c$} 
\put(50,20){$1$} 
\put(70,20){$1'$} 
\put(85,3){$c$} 
\put(30,0){\line(1,0){60}} 
\put(50,0){\line(0,1){15}} 
\put(70,0){\line(0,1){15}} 
\end{picture}
\label{eqn:T-defn} 
\end{equation} 
where $1$ represents the spacetime location 
$\theta=\pi/2$, $\tau=\tau_1$, while $1'$ represents the spacetime 
location $\theta=3\pi/2$, $\tau=\tau_1$.  The index $c$ can be either 
$I$ or $\psi$; the  fact that $a$ is the 
same on both sides here is the precise meaning of ${\cal T}_\sigma$ being 
in the identity channel. 
 
The tunneling Lagrangian for a point contact in a $p+ip$ 
superconductor is therefore 
$${\cal L}_{\hbox{tun}} = \lambda_\psi 
{\cal T}_\psi(\tau) + \lambda_{\sigma} {\cal T}_\sigma(\tau)$$
with ${\cal T}_\sigma(\tau)$ defined as in (\ref{eqn:T-defn}). 
The partition function is then defined perturbatively by  
expanding in powers of $\lambda_\sigma$ and 
$\lambda_\psi$. The coefficients are the conformal blocks  
defined by the requirement that 
${\cal T}_\sigma$ be in the identity channel. Such   
conformal blocks of ${\cal T}_\sigma$ then have all 
$c_j=I$, i.e.\ $\langle{\cal T}_\sigma(\tau_1){\cal T}_\sigma(\tau_2) 
\dots {\cal T}_\sigma(\tau_{j})\rangle$ is pictorially represented as 
\begin{center} 
\begin{picture}(250,30) 
\put(13,3){$I$} 
\put(30,20){$1$} 
\put(50,20){$1'$} 
\put(90,20){$2$} 
\put(110,20){$2'$} 
\put(70,3){$I$} 
\put(10,0){\line(1,0){125}} 
\put(30,0){\line(0,1){15}} 
\put(50,0){\line(0,1){15}} 
\put(90,0){\line(0,1){15}} 
\put(110,0){\line(0,1){15}} 
\put(125,3){$I$} 
\put(145,3){$\dots$} 
\put(165,3){$I$} 
\put(165,0){\line(1,0){65}} 
\put(190,0){\line(0,1){15}} 
\put(210,0){\line(0,1){15}} 
\put(190,20){$j$} 
\put(210,20){$j'$} 
\put(227,3){$I$} 
\end{picture} 
.\end{center}

To make this more concrete, we give the simplest cases explicitly. 
There is only one vanishing two-point function of two 
sigma operators, which on the plane is 
$$ 
\begin{picture}(100,30) 
\put(13,3){$I$} 
\put(30,20){$1$} 
\put(50,20){$2$} 
\put(10,0){\line(1,0){60}} 
\put(30,0){\line(0,1){15}} 
\put(50,0){\line(0,1){15}} 
\put(67,3){$I$} 
\end{picture} 
=\frac{1}{(z_1-z_2)^{1/8}}.$$ 
The four-point conformal blocks ${\cal F}_c$ discussed above were defined by  
demanding that the fields at $z_1$ and $z_2$ fuse to the $c$ channel, so 
\begin{center} 
\begin{picture}(140,30) 
\put(-20,3){${\cal F}_c\equiv$} 
\put(13,3){$I$} 
\put(30,20){$1$} 
\put(50,20){$2$} 
\put(90,20){$3$} 
\put(110,20){$4$} 
\put(70,3){$c$} 
\put(10,0){\line(1,0){125}} 
\put(30,0){\line(0,1){15}} 
\put(50,0){\line(0,1){15}} 
\put(90,0){\line(0,1){15}} 
\put(110,0){\line(0,1){15}} 
\put(125,3){$I$} 
\end{picture} 
. 
\end{center} 
Explicitly, one finds \cite{Belavin84}  
\begin{eqnarray} 
\nonumber 
{\cal F}_I &=& \left(\frac{1}{z_{12}z_{34}x(1-x)}\right)^{1/8} 
\left(1+\sqrt{1-x}\right)^{1/2}, \\ 
{\cal F}_\psi &=& \left(\frac{1}{z_{12}z_{34}x(1-x)}\right)^{1/8} 
\left(1-\sqrt{1-x}\right)^{1/2},  
\label{fourpointexplicit} 
\end{eqnarray} 
where $z_{ij}=z_i-z_j$ and $x=z_{12}z_{34}/z_{13}z_{24}$.  
 
More complicated conformal blocks can be 
computed explicitly by using the algebraic structure of the conformal 
field theory to derive differential equations for them; specifying the 
fusion channels for a given conformal block then amounts to 
choosing the particular solution of the differential 
equation. Computing conformal blocks in this fashion gets
very tedious beyond low 
orders. In section \ref{sec:bosonization}
we therefore explain how to compute them 
using a much simpler procedure: bosonization.

\section{Cluster Decomposition of Conformal Blocks} 
\label{sec:decomposition}

In the previous section, we showed how to properly define the quasiparticle
tunneling operator. One notable feature
of the definition is that a quasiparticle on one side of
the point contact will be entangled with the quasihole
which was left behind on the other side when it tunneled across.
This is necessitated by the physics of non-Abelian
statistics. However, this kind of non-local correlation
makes it difficult to treat the two edges as independent. Nonetheless,
with a little work we can disentangle the fields on the two edges,
enabling us in the next section to bosonize the quasiparticle
tunneling operator.

It should be possible to treat the two $\sigma$ fields in ${\cal
T}_\sigma$ as independent when the circumference of the disc is
``long'', meaning that the radius $R$ of the disk is much larger than
the inverse temperature $\beta$.  In this limit, conformal blocks of
tunneling operators obey cluster decomposition: correlations between
fields on the opposite sides fall off as some power of $\beta/R$
relative to correlations of fields at the same spatial point but at
different times.  The tension here is that we would like to treat
$\sigma(\pi/2)$ and $\sigma(3\pi/2)$ as two independent fields,
$\sigma_a$ and $\sigma_b$, since the points $\theta=\pi/2$ and
$\theta=3\pi/2$ are far apart so their correlations satisfy cluster
decomposition.  On the other hand, the tunneling operator is defined
by the fusion channel of $\sigma(\pi/2)$ and $\sigma(3\pi/2)$.
Although $\theta=\pi/2$ and $\theta=3\pi/2$ are far apart, the choice
of fusion channel is topological and is insensitive to the distance
between these points. Since $\sigma(\pi/2)$ and $\sigma(3\pi/2)$ are
entangled in this way, we cannot cluster decompose their correlation
functions even though $\theta=\pi/2$ and $\theta=3\pi/2$ are far
apart.

The resolution is to switch into a basis in which we specify the fusion
of $\sigma$ fields at the same spatial point. Consider the four-point
conformal blocks. We can define an alternate basis
${\cal G}_c$, for this two-dimensional vector space: 
$${\cal G}_c\equiv 
\begin{picture}(130,30) 
\put(13,3){$I$} 
\put(30,20){$1$} 
\put(50,20){$2$} 
\put(90,20){$1'$} 
\put(110,20){$2'$} 
\put(70,3){$c$} 
\put(10,0){\line(1,0){125}} 
\put(30,0){\line(0,1){15}} 
\put(50,0){\line(0,1){15}} 
\put(90,0){\line(0,1){15}} 
\put(110,0){\line(0,1){15}} 
\put(125,3){$I$} 
\end{picture} 
$$ with $c=I,\psi$. In this diagram, the number $j$ represents the
spacetime location ($\theta=\pi/2$, $\tau=\tau_j$), while $j'$
represents the spacetime location ($\theta=3\pi/2$, $\tau=\tau_j$).
Both bases, ${\cal F}_c$ and ${\cal G}_c$, are perfectly valid. The
advantage of the ${\cal F}_c$ basis is that it arises naturally in
perturbation theory.  The advantage of the ${\cal G}_c$ basis is that
$\sigma(\pi/2)$ and $\sigma(3\pi/2)$ are not entangled with each
other. Cluster decomposition then gives ${\cal G}_I$ and ${\cal
G}_\psi$ each to be the product of two-point functions for each edge
(to leading order in $\beta/R$). This means that ${\cal G}_\psi=0$ to
this order, because it doesn't have $I$ on both ends.  Neglecting
terms suppressed by powers of $\beta/R$, we have
\begin{eqnarray} 
\nonumber 
{\cal G}_I &=&  
\begin{picture}(150,30) 
\put(3,3){$I$} 
\put(20,20){$1$} 
\put(40,20){$2$} 
\put(0,0){\line(1,0){60}} 
\put(20,0){\line(0,1){15}} 
\put(40,0){\line(0,1){15}} 
\put(57,3){$I$} 
\put(83,3){$I$} 
\put(100,20){$1'$} 
\put(120,20){$2'$} 
\put(80,0){\line(1,0){60}} 
\put(100,0){\line(0,1){15}} 
\put(120,0){\line(0,1){15}} 
\put(137,3){$I$} 
\end{picture} 
\\ 
&=& 
\left(\frac{1}{\sin(\tau_1-\tau_2)}\right)^{1/4} 
\label{GI} 
\end{eqnarray} 
To obtain the latter we conformally map the punctured plane to a 
infinitely-long cylinder by using $w=\exp(\theta R/\beta+i\tau)$.  
These results for ${\cal G}_I$ and ${\cal G}_\psi$ can easily be 
checked by taking $R\gg\beta$ limit of the full four-point conformal 
blocks, which were originally derived in ref. \onlinecite{Belavin84}.

When the $2n$-point correlation function is decomposed into
the product of the $n$-point correlation function of $\sigma_a$s
multiplied by the $n$-point correlation function of $\sigma_b$s,
we must then consider the conformal blocks for each of
these. In ref. \onlinecite{Fendley06a}, we introduced
the notation $({m_1},\ldots,m_{n/2})_{a(b)}$ with ${m_i}=0,1$
to specify the conformal blocks on the a(b) edges, respectively. 
In terms of the ``vertex" notation described above, 
\begin{multline}
({m_1},\ldots,m_{n/2})_{a(b)} =\\ 
\begin{picture}(250,30) 
\put(10,3){$I$} 
\put(30,20){$1$} 
\put(50,20){$2$} 
\put(90,20){$3$} 
\put(110,20){$4$} 
\put(67,3){$c_1$} 
\put(10,0){\line(1,0){125}} 
\put(30,0){\line(0,1){15}} 
\put(50,0){\line(0,1){15}} 
\put(90,0){\line(0,1){15}} 
\put(110,0){\line(0,1){15}} 
\put(125,3){$c_2$} 
\put(145,3){$\dots$} 
\put(165,5){$c_{\frac{n}{2}-1}$} 
\put(165,0){\line(1,0){65}} 
\put(190,0){\line(0,1){15}} 
\put(210,0){\line(0,1){15}} 
\put(178,20){$n-1$} 
\put(210,20){$n$} 
\put(227,3){$I$} 
\end{picture}\\
\label{eqn:parenthesis-notation}
\end{multline}
with $c_j=I$ or $\psi$ if $\sum_{i=1}^{j}{m_i}$ is $0 (\text{mod} 2)$
or $1 (\text{mod} 2)$ respectively.
In this notation, ${{\cal G}_I}={(0)_a}\cdot{(0)_b}$,
and ${{\cal G}_\psi}={(1)_a}\cdot{(1)_b}=0$.

In order to make use of the ${\cal G}_c$ basis, we must
transform to it from the ${\cal F}_c$ basis which arises in
perturbation theory for the tunneling operator. Consider 
$$\langle {\cal T}_\sigma(\tau_1) {\cal T}_\sigma(\tau_2)\rangle= 
\begin{picture}(130,30) 
\put(13,3){$I$} 
\put(30,20){$1$} 
\put(50,20){$1'$} 
\put(90,20){$2$} 
\put(110,20){$2'$} 
\put(70,3){$I$} 
\put(10,0){\line(1,0){125}} 
\put(30,0){\line(0,1){15}} 
\put(50,0){\line(0,1){15}} 
\put(90,0){\line(0,1){15}} 
\put(110,0){\line(0,1){15}} 
\put(125,3){$I$} 
\end{picture} 
$$ 
which is simply ${\cal F}_I$. As Moore and Seiberg\cite{Moore88}
explained in an in-depth analysis of the topological
properties of conformal blocks, there is a linear transformation
which connects this to the ${\cal G}_c$ basis:
\begin{equation}
{\cal F}_c = \sum_d {\cal B}_{cd}{\cal G}_d
\end{equation}
Moore and Seiberg\cite{Moore88} showed that
the requirement that one obtains the same non-chiral 
correlators from any basis for conformal blocks
results in a huge number of constraints on the matrix ${\cal B}_{cd}$.
This matrix is called the ``braid matrix'', for fairly obvious reasons:
the only difference between ${\cal F}$ and ${\cal G}$ is the
exchange or braid of $2$ with $1'$.  
For the Ising model, the braid matrix for two $\sigma$ fields with $I$ 
or $\psi$ (such as, for instance, $1'$ and $2$) connecting them is 
\cite{Moore88} 
\begin{equation}
{\cal B} =  
\frac{e^{-i\pi/8}}{\sqrt{2}} 
\begin{pmatrix} 
1&i\\ 
i&1 
\end{pmatrix} 
\label{braidIsing} 
\end{equation}
This braid matrix can be applied not only to the
conformal blocks (\ref{fourpointexplicit}), but to
any conformal block (tensored by the identity acting
on all other indices).

Here, we are using a braiding operation in a
`passive' sense: not as an actual physical braid,
but as a change of basis. Indeed, any element of the braid
group can be viewed either in an `active' sense
(moving particles around) or in a `passive' sense
of a change of basis.
The braid matrices associated with this chiral conformal field
theory are the same as those associated with the
Chern-Simons topological field theory for the bulk state \cite{Witten89}.
In the latter context, it is more natural to view the braid
in the `active' sense: as quasiparticles are exchanged in
the bulk, different topologically degenerate states are
transformed into each other. The elements of the braid matrices
can be viewed as the amplitudes for various quasiparticle
histories, which can be computed from
the Jones-Kauffman invariants
of knot theory \cite{Jones,Kauffman}.
For any topological class of quasiparticle trajectories,
we can thereby associate a braid matrix.

Therein lies a quandary: what braiding
operation should we use? If we simply use ${\cal B}$, then
we obtain 
\begin{equation} 
\langle{\cal T}(\tau_1){\cal T}(\tau_2) \rangle = {\cal F}_I =e^{i\pi/8} ({\cal 
G}_I + i {\cal G}_\psi). 
\label{FG} 
\end{equation} 
However, we could just as well use ${\cal B}^{-1}$,
which exchanges $1'$ and $2$ in a clockwise (rather
than a counterclockwise) manner. Indeed, we could then
follow this with any braid which is diagonal in the ${\cal G}_a$ 
basis, such as winding $1'$ around $2'$ any number of times.
So which braid (interpreted in a passive sense) should
we use? Said differently, there are many possible bases
${\cal G}_c$ which are diagonal in the fusion $1$, $2$
and $1'$, $2'$ fusion channels. Which of these should we
use to decompose the $\sigma$ multi-point correlation functions?

\begin{figure}[t!]
\centerline{\includegraphics[width=3.05in]{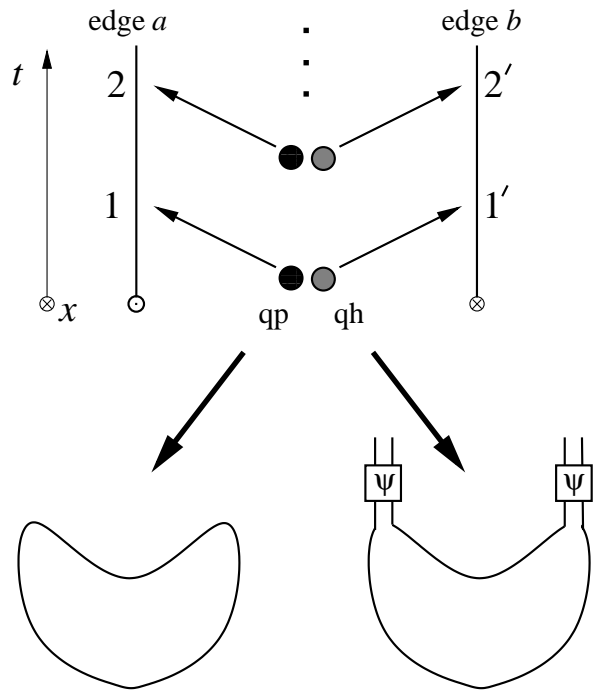}}
\caption{Spacetime history of quasiparticle-quasihole pair creation
processes contributing to transport across the point contact.
A pair is created at the middle of the point contact; the quasiparticle
moves to one edge and the quasihole moves to the other.
When two such processes happen in succession, we can ask how
the two quasiparticles which end up on the same edge fuse.
The amplitude for such fusion into the $1$ and $\psi$ channels
is given by the Jones-Kauffman bracket evaluation of the knot diagrams
at the bottom left and right
of the figure.}
\label{fig:tunneling-process}
\end{figure}

To answer this question, let us take a closer look
at the tunneling process. We can view it as the creation
of a quasiparticle-quasihole pair at the middle of the junction
and the subsequent motion of the quasiparticle to one
edge at $1$ and the quasihole to the other edge at $1'$,
as depicted in figure \ref{fig:tunneling-process}.
Since the quasiparticle-quasihole
pair is created out of the ground state, it fuses to the identity,
as we argued in the previous section. The next tunneling process
occurs in the same way: a quasiparticle-quasihole pair (which
fuses to the identity) is created
from the ground state, and the quasiparticle and quasihole move
to the two edges at $2$ and $2'$. We would now like to know how
$1$ and $2$ fuse and how $1'$ and $2'$ fuse. At the bottom
of figure \ref{fig:tunneling-process}, we have redrawn the spacetime
history of the four quasiparticles in the form of knot diagrams
which specify the desired basis change. 

\begin{figure}[t!]
\centerline{\includegraphics[width=3.25in]{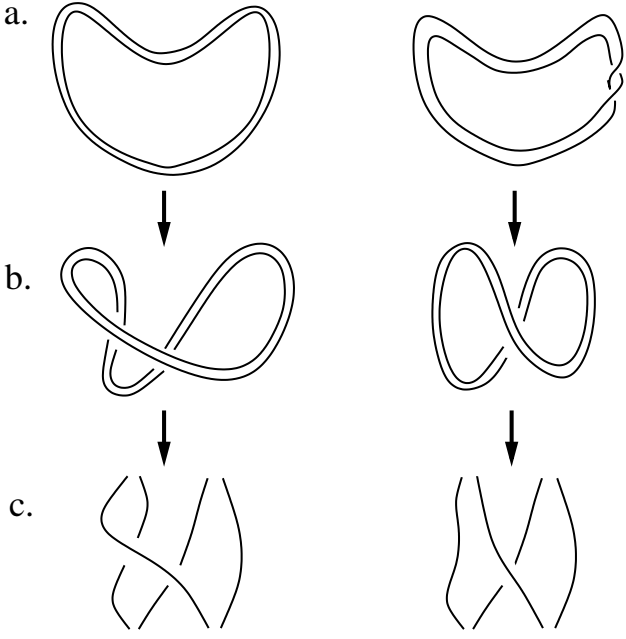}}
\caption{(a) Two different histories in which two quasiparticle
quasiholes are created and the two quasiparticles are fused
and the two quasiholes are fused. In the history on the right,
one of the quasiparticles twists by $2\pi$ before fusing. (b) A topologically
equivalent set of histories in which the pair creations and annihilations
occur at the same times. (c) The braid group elements corresponding to
these histories.}
\label{fig:fusion-braid}
\end{figure}

However, there is still an ambiguity in the change of basis.  Because
the quasiparticles are created by chiral fields, they are not
invariant under a rotation by $2\pi$: they pick up phases, just like a
fermion picks up a minus sign. To keep track of these phases, we must
give a framing to these histories. This is done pictorially by thickening
these lines into ribbons. This is physically natural since
quasiparticles have a finite size. It is also mathematically essential;
otherwise, the distinction between different braids will be lost. For
instance, without the framing, the two pictures in figure
\ref{fig:fusion-braid}b will be topologically equivalent, even though
they are associated with different braids in \ref{fig:fusion-braid}c.
The framing is determined by the physics of the situation. In this
case, the geometry of the tunnel junction prefers the `blackboard
framing', in which the curve in figure \ref{fig:fusion-braid}a is
thickened into a ribbon which is contained entirely in the plane of
the page (or the proverbial blackboard on which it is drawn). In any
other history, one of the quasiparticles would have to spin around, as
in the picture on the right side of figure \ref{fig:fusion-braid}a,
which would be energetically costly.

In general, when we have four operators in 
a row, i.e.\ 
\begin{center} 
\begin{picture}(160,30) 
\put(-20,3){$\dots$} 
\put(10,3){$I$} 
\put(15,20){$2j$--$1$} 
\put(40,20){$(2j$--$1)'$} 
\put(85,20){$2j$} 
\put(105,20){$(2j)'$} 
\put(70,3){$I$} 
\put(10,0){\line(1,0){125}} 
\put(30,0){\line(0,1){15}} 
\put(50,0){\line(0,1){15}} 
\put(90,0){\line(0,1){15}} 
\put(110,0){\line(0,1){15}} 
\put(125,3){$I$} 
\put(145,3){$\dots$} 
\end{picture} 
\end{center} 
we braid the operators at $(2j-1)'$ and $2j$ and then
braid $(2j-1)'$ with $(2j)'$. We then obtain the ordering 
\begin{equation} 
\begin{picture}(160,30) 
\put(-20,3){$\dots$} 
\put(10,3){$I$} 
\put(15,20){$2j$--$1$} 
\put(103,20){$(2j$--$1)'$} 
\put(42,20){$2j$} 
\put(80,20){$(2j)'$} 
\put(70,3){$c$} 
\put(10,0){\line(1,0){125}} 
\put(30,0){\line(0,1){15}} 
\put(50,0){\line(0,1){15}} 
\put(90,0){\line(0,1){15}} 
\put(110,0){\line(0,1){15}} 
\put(125,3){$I$} 
\put(145,3){$\dots$} 
\end{picture}. 
\label{TT} 
\end{equation} 
To determine $c$, we need the relation \cite{Moore88} 
\begin{equation} 
\begin{picture}(250,30) 
\put(20,3){$c$} 
\put(20,0){\line(1,0){40}} 
\put(40,0){\line(0,1){15}} 
\put(40,20){$a$} 
\put(57,3){$b$} 
\put(73,5){$= e^{i\pi(h_a+h_b-h_c)}$} 
\put(150,0){\line(1,0){40}} 
\put(150,3){$c$} 
\put(170,20){$b$} 
\put(170,0){\line(0,1){15}} 
\put(187,3){$a$} 
\end{picture} 
\label{threepoint} 
\end{equation}
where $h_a$ is the chiral scaling dimension of the operator labelled
by $a$. (For more complicated theories than the Ising model, there can
be extra signs in (\ref{threepoint}), but they do not appear here.) We
have already seen that $h_\sigma=1/16$ and $h_\psi=1/2$. Putting this
all together yields $c=I+\psi$, i.e.\cite{Moore88}
\begin{multline} 
\langle \dots {\cal  
T}_\sigma(\tau_{2j-1}) {\cal T}_\sigma(\tau_{2j})\dots\rangle\\ 
= 
\begin{picture}(170,30) 
\put(0,3){$\dots$} 
\put(30,3){$I$} 
\put(35,20){$2j$--$1$} 
\put(125,20){$(2j$--$1)'$} 
\put(65,20){$2j$} 
\put(102,20){$(2j)'$} 
\put(82,3){$I+\psi$} 
\put(30,0){\line(1,0){125}} 
\put(50,0){\line(0,1){15}} 
\put(70,0){\line(0,1){15}} 
\put(110,0){\line(0,1){15}} 
\put(130,0){\line(0,1){15}} 
\put(145,3){$I$} 
\put(165,3){$\dots$} 
\end{picture} 
\label{TT2} 
\end{multline} 
Recall that conformal blocks are elements of a vector space, so 
$I+\psi$ just means the sum of the blocks with a $I$ and a $\psi$ in the middle. 
Using this braiding, we can now unambiguously decompose any of our 
conformal blocks into the product of conformal blocks on the two 
edges.

\section{Bosonizing conformal blocks}
\label{sec:bosonization} 
 
We now describe a method which not only allows us
to compute conformal blocks explicitly, but also to
write ${\cal T}_\sigma$ in a form in which 
the physics is much clearer. This is to use {\em bosonization} to 
write ${\cal T}_\sigma$ in terms of a free boson $\phi_\sigma$. (This is not 
to be confused with the charge boson $\phi_\rho$ associated with the 
Moore-Read state, but rather a new boson introduced to make 
computation of the conformal blocks possible.) 
 
It has long been known how to write the correlation functions
of the non-chiral Ising model in terms of the correlation functions
of a free boson \cite{DSZboso}. The 
trick is to note that one can combine two independent Majorana 
fermions into a single Dirac fermion, which can be bosonized.
The current is bosonized as:
\begin{equation}
\label{eqn:current-bosonized}
i{\psi_a}{\psi_b} = \frac{1}{2\pi}\,{\partial_x}{\phi_\sigma}.
\end{equation}
Meanwhile, the Dirac fermion
\begin{equation}
{\psi_a}+i{\psi_b} = e^{i\phi_\sigma}.
\label{bosonization-def}
\end{equation}
In order to use this method, we need (1) to treat
the two edges as independent and (2) to use the `flipped' setup;
since both of the Majorana modes then have the same chirality,
they can be combined into a Dirac fermion.

However, representing the chiral spin field, $\sigma(z)$,
is trickier. The full non-chiral spin fields
${\sigma_a}(z,\overline{z})$ and ${\sigma_b}(z,\overline{z})$
can be bosonized according to \cite{DSZboso} 
\begin{equation} 
{\sigma_a}(z,\overline{z}){\sigma_b}(z,\overline{z}) =  
\sqrt{2}\cos(({\phi_\sigma}(z)+\overline{{\phi_\sigma}}(\overline{z}))/2) 
\label{DSZboso} 
\end{equation}
where we have now introduced $\overline{{\phi_\sigma}}(\overline{z})$,
which is a left-moving counterpart to ${\phi_\sigma}(z)$.  The square
of any non-chiral Ising correlator can be written as a product of a
correlator of fields with label $a$ with a correlator with $b$ labels,
since the $a$ and $b$ fields are independent. Since correlators
of exponentials of a boson are easily evaluated,
the correlators of $\cos(({\phi_\sigma}+\overline{{\phi_\sigma}})/2)$
are also easy to find, yielding the Ising correlator as their square
root.

We now show how to extend (\ref{DSZboso}) to write the conformal 
blocks of the tunneling operator ${\cal T}_\sigma$ in terms of 
bosons. The formula (\ref{DSZboso}) does not decompose
into a product of chiralities, so 
it takes more work to find bosonic expressions for chiral sigma 
fields. One method is to use the fact that the
product of {\em dis}order fields has a bosonic expression
\begin{equation} 
{\mu_a}(z,\overline{z}){\mu_b}(z,\overline{z}) =  
\sqrt{2}\sin(({\phi_\sigma}(z)+\overline{{\phi_\sigma}}(\overline{z}))/2) 
\end{equation}
By using the operator product expansion, one can derive bosonized 
expressions for ${\sigma_a}{\mu_b}$ and ${\mu_a}{\sigma_b}$ as well. 
Since both $\sigma_a$ and $\mu_a$ can be expressed in
terms of chiral and antichiral 
vertex operators, one can then find linear combinations which split into 
the product of chiral and antichiral vertex operators.
This is conceptually straightforward but 
in practice is tricky, because one must keep track of a variety of 
phases. It is much simpler to deal with the chiral conformal blocks 
directly. The bosonization of the non-chiral fields does give us 
valuable information, namely that bosonization of conformal blocks is 
possible. Consider the product
${{\cal T}_\sigma}({z_1})\,{{\cal T}_\sigma}({z_2})$, as in (\ref{TT}). We 
have a four-dimensional space of conformal blocks, corresponding to 
having either $I$ or $\psi$ in the middle and at the right end. All 
correlators of the non-chiral spin and disorder fields can be built up 
by products of these blocks. As discussed earlier, we then take 
the limit of a ``long'' disc, so that the fields on opposite sides can 
be treated independently. In the bosonized formulation, there are four 
linearly-independent chiral fields describing these conformal blocks, 
namely 
\begin{equation} 
e^{\pm i{\phi_\sigma}(z_1)/2}e^{\pm i{\phi_\sigma}(z_2)/2} 
\label{pmpm} 
\end{equation} 
for any of the four choices of $\pm$ signs. Since both spaces are 
four-dimensional, and the elements of each are linearly independent, 
we must be able to write one in terms of the other.  
 
Since we know that products of tunneling operators in (\ref{TT2}) can 
be bosonized, consistency under braiding allows us to find 
exactly what they are. One key fact is that braiding operations leave 
the overall fusion channel for
${\cal T}_\sigma(\tau_{2j-1}){\cal T}_\sigma(\tau_{2j})$
unchanged, i.e. the ends of 
(\ref{TT}) and (\ref{TT2}) remain $I$ no matter what braiding we do 
inside. The bosonized version must do likewise, so only two of the 
four fields in (\ref{pmpm}) can contribute to (\ref{TT}) or 
(\ref{TT2}),  
namely 
$$e^{i{\phi_\sigma}(z_1)/2}e^{- i{\phi_\sigma}(z_2)/2}\quad\hbox{ and }\quad 
e^{-i{\phi_\sigma}(z_1)/2}e^{i{\phi_\sigma}(z_2)/2}.$$ 
Hence, we find that 
\begin{multline} 
\begin{picture}(170,30) 
\put(0,3){$\dots$} 
\put(30,3){$I$} 
\put(35,20){$2j$--$1$} 
\put(123,20){$(2j$--$1)'$} 
\put(62,20){$2j$} 
\put(100,20){$(2j)'$} 
\put(90,3){$c$} 
\put(20,0){\line(1,0){125}} 
\put(50,0){\line(0,1){15}} 
\put(70,0){\line(0,1){15}} 
\put(110,0){\line(0,1){15}} 
\put(130,0){\line(0,1){15}} 
\put(145,3){$I$} 
\put(165,3){$\dots$} 
\end{picture}\\ 
\propto \dots\left(e^{i{\phi_\sigma}(z_1)/2}e^{- i{\phi_\sigma}(z_2)/2} \pm e^{- 
i{\phi_\sigma}(z_1)/2}e^{i{\phi_\sigma}(z_2)/2}\right)\dots 
\label{cossin} 
\end{multline} 
where the relative $+$ sign is for $c=I$, and the minus sign for 
$c=\psi$. The simplest way of seeing this is to note that in the limit of 
a long disk, non-vanishing conformal blocks must contain an even number of 
pieces like (\ref{cossin}) with $c=\psi$. For example,  
we saw above that the four-point function ${\cal G}_\psi=0$ in this 
limit. 
Since correlators of free bosons are invariant under sending ${\phi_\sigma}$ to 
$-{\phi_\sigma}$, having ${\cal G}_\psi=0$ requires the $-$ sign for $c=\psi$, which 
then requires the $+$ sign for $c=I$.
We can restate this result in the notation
of (\ref{eqn:parenthesis-notation}): 
\begin{multline}
\label{eqn:blocks-bosonized}
{\left({m_1},{m_2},\ldots,{m_{n/2}}\right)^2} \Bigl\langle \prod_{j=1}^{n/2} 
\Big(e^{i\left({\phi_\sigma}({\tau^{}_{2j-1}})
-{\phi_\sigma}({\tau^{}_{2j}})\right)/2}\\
+(-1)^{m_j}
e^{-i\left({\phi_\sigma}({\tau^{}_{2j-1}})
-{\phi_\sigma}({\tau^{}_{2j}})\right)/2} \Big)\Bigr\rangle
\end{multline} 
 
As a consistency check on (\ref{cossin}), consider interchanging $1$ 
and $2$, and $2$ with $1'$. (This amounts to removing the 
aforementioned ambiguity by changing $1$ with $2$ instead of $2'$ with 
$1'$.)  This interchange results in an overall phase $e^{i\pi/4}$ for 
$c=I$, while $c=\psi$ gets a phase $-e^{i\pi/4}$, the extra minus sign 
coming from the two factors of $i$ in (\ref{threepoint}).  In the bosonic 
formulation, this interchange just corresponds to braiding the two 
$e^{\pm i{\phi_\sigma}}$. The operator product expansion yields
$$e^{i{\phi_\sigma}(z_1)/2}e^{- i{\phi_\sigma}(z_2)/2} = e^{i\pi/4}
e^{- i{\phi_\sigma}(z_2)/2}e^{i{\phi_\sigma}(z_1)/2},$$ giving the
needed factor of $e^{i\pi/4}$. The extra minus sign for $c=\psi$ then
arises because the two terms on the right-hand-side of (\ref{cossin})
have been interchanged.
 
The last (and trickiest) thing to get straight is the phase in front 
of (\ref{cossin}). The phase for $c=I$ can be fixed 
by comparing to ${\cal G}_I$, but the important part is the relative 
phase between $c=I$ and $c=\psi$. The importance is because as shown 
in (\ref{TT2}), the conformal blocks contain the sum of the two terms. 
To get this sign straight, consider a conformal block of the form 
$(\ref{TT})$ with $N_\psi$ intermediate $\psi$ states. This means we 
have $N_\psi$ primed states in the $\psi$ channel, and $N_\psi$ 
umprimed states in the $\psi$ channel. To cluster 
decompose the correlator, we need to move all the primed fields 
together at one end, and the unprimed fields to the other. This is 
done by the usual braiding rules, and does not change the block, 
except when a pair of fields in the $\psi$ channel is interchanged 
with another pair in the $\psi$ channel, where one picks up a minus 
sign. Moving all the fields to the appropriate sides results in an 
overall sign $(-1)^{N_\psi/2}$ ($N_\psi$ must be even for the 
conformal block to not vanish in the long-disc limit.) However, this 
sign is cancelled by the fact that to get the same conformal blocks 
for the primed and unprimed cases, we need to braid all the 
$(2j)'$ and $(2j-1)'$ terms, so that the primed block will be ordered 
$1'2'3'4'\dots$ instead of $2'1'4'3'\dots$. This gives us a factor  
$i^{N_\psi}=(-1)^{N_\psi/2}$, cancelling the previous factor. 
 
The upshot is that the two cases in (\ref{cossin}) have the same 
overall phase, which can be absorbed in the coefficient 
$\lambda_\sigma$.  
We have now bosonized both basis elements for the product of tunneling 
operators. To obtain the tunneling operator, we now just add the
bosonized expression for $I$ and $\psi$ channels together, as
indicated in (\ref{TT2}). Our result is therefore
\begin{equation}
\label{eqn:tunneling-block} 
{\cal T}_\sigma(\tau_{2j-1}){\cal T}_\sigma(\tau_{2j})=  
e^{i{\phi_\sigma}(\tau_{2j-1})/2}e^{- i{\phi_\sigma}(\tau_{2j})/2} 
\end{equation} 
Thus the conformal blocks in the perturbative expansion in powers of 
$\lambda_\sigma$ are given by a single correlator of bosonic vertex 
operators. These are easily evaluated, with $2n$ tunneling operators 
we have 
$$ \left|\frac{\prod_{i<j\le n}  
\sin((\tau_{2i-1}-\tau_{2j-1})/2) \sin((\tau_{2i}-\tau_{2j})/2)} 
{\prod_{i\le j\le n}  
\sin((\tau_{2i-1}-\tau_{2j})/2)}\right|^{1/4}  
$$

With these bosonization formulas in hand, we can compute
all of the conformal blocks of the Ising model, not just the combination
(\ref{eqn:tunneling-block}). See appendix \ref{sec:block-computation}
for details.

\section{Mapping to the Kondo problem and Resonant Tunneling
between Luttinger Liquids}
\label{sec:Kondo}

In the previous section, we learned that the perturbation
expansion of the partition function and correlation functions
can be bosonized using:
\begin{multline}
\left\langle \ldots
{\cal T}_\sigma(\tau_1){\cal T}_\sigma(\tau_{2})\ldots
{\cal T}_\sigma(\tau_{2k-1}){\cal T}_\sigma(\tau_{2k})
\right\rangle= \\ 
\left\langle \ldots e^{i{\phi_\sigma}(\tau_{1})/2}e^{- i{\phi_\sigma}(\tau_{2})/2}
\ldots e^{i{\phi_\sigma}(\tau_{2k-1})/2}e^{- i{\phi_\sigma}(\tau_{2k})/2}\right\rangle
\end{multline}
On the right-hand-side, the operators $e^{i{\phi_\sigma}}$
and $e^{- i{\phi_\sigma}}$ alternate. This is precisely the same as
$$
\left\langle\left({S^+}e^{-i{{\phi_\sigma}}/2}+{S^-}e^{i{{\phi_\sigma}}/2}\right)
\left({S^+}e^{-i{{\phi_\sigma}}/2}+{S^-}e^{i{{\phi_\sigma}}/2}\right)
\ldots \right\rangle
$$
if $\vec{S}$ is a spin-$1/2$ operator. Since
${\left({S^+}\right)^2}={\left({S^+}\right)^2}=0$,
$e^{i{\phi_\sigma}}$ and $e^{- i{\phi_\sigma}}$ alternate
in the same way in this expression, too.
 
Therefore, we can rewrite the Hamiltonian (\ref{eqn:p+ip-action})
for a point contact in a $p+ip$ superconductor
in the bosonic form:
\begin{multline}
\label{eqn:bosonic-p+ip}
{\cal H}_{p+ip} =  \int dx\left(\frac{v_n}{2\pi}{\left({\partial_x}{\phi_\sigma}\right)^2}\right)
\: + \: \frac{\lambda_\psi}{2\pi} {\partial_x}{\phi_\sigma}{\hskip -0.03cm}(0)\\
+ {\lambda_\sigma}\left({S^+}e^{-i{\phi_\sigma}{\hskip -0.03cm}(0)/2} +
{S^-}e^{i{\phi_\sigma}{\hskip -0.03cm}(0)/2}\right)
\end{multline}
The Hamiltonian for a Moore-Read Pfaffian state at $\nu=5/2$
can be bosonized similarly. The charge $e/4$ quasiparticle
tunneling operator is the product of
${\cal T}_\sigma=\left({S^+}e^{-i{\phi_\sigma}{\hskip -0.03cm}(0)/2}+
{S^-}e^{i{\phi_\sigma}{\hskip -0.03cm}(0)/2}\right)$ with
the cosine of the charged mode, $\cos\!\left({\phi_\rho}(0)/2\right)$.
The neutral fermion tunneling operator
is the same as in (\ref{eqn:bosonic-p+ip}). The charge $e/2$ tunneling
operator is independent of the Majorana fermion and is the same as
in (\ref{eqn:flipped-five-halves-action}). Hence, we have the Hamiltonian
\begin{multline}
\label{eqn:bosonic-Pfaffian}
{\cal H}_{5/2} = \int dx\left(\frac{v_c}{2\pi}\left({\partial_x}{\phi_\rho}\right)^2
+\frac{v_n}{2\pi}\left({\partial_x}{\phi_\sigma}\right)^2\right)\\
+  \lambda_{1/4} \left({S^+}e^{-i{\phi_\sigma}{\hskip -0.03cm}(0)/2}+
{S^-}e^{i{\phi_\sigma}{\hskip -0.03cm}(0)/2}\right) \cos\!\left({\phi_\rho}(0)/2\right)\\
+ \: \lambda_{1/2} \cos{\phi_\rho}(0) \: +
\: \frac{\lambda_{\psi,0}}{2\pi}\, {\partial_x}{\phi_\sigma}{\hskip -0.03cm}(0),
\end{multline}
An advantage of the bosonic formulation is that it allows
for a semiclassical analysis in the infrared limit
and it reveals the relationship between our problem and
resonant tunneling between
Luttinger liquids \cite{Kane92_Resonant}, the Kondo problem \cite{Emery92},
and dissipative quantum mechanics \cite{Leggett87}.

The bosonic Hamiltonian for a $p+ip$
superconductor (\ref{eqn:bosonic-p+ip}):
is literally the single-channel Kondo problem.
To see this, consider the anisotropic Kondo Hamiltonian
\begin{multline}
\label{eq:Kondo}
{\cal H}_{\rm Kondo} = {\cal H}_{\rm cond} + {J_z}\,{S^z}\cdot{s^z}(0)\\
+ {J_\perp}\, \left({S^+}\cdot{s^-}(0) + {S^-}\cdot{s^+}(0)\right)
\end{multline}
where ${\bf S}$ is the impurity spin and
${\bf s}({\bf x})={\psi_\alpha^\dagger}\vec{\sigma}_{\alpha\beta}\psi_\beta$
is the conduction electron spin density. Since the impurity spin
only interacts with electrons in the $s$-wave channel about the
origin, we can focus on this channel and treat the problem
as one-dimensional. If there is only a single channel of
conduction electrons, then when we bosonize the $s$-wave
electrons, we obtain:
\begin{multline}
\label{eqn:1-ch-Kondo-boson}
{\cal H}_\text{1-ch. Kondo} = \int dx\,\frac{v_F}{2\pi}
\left(\left({\partial_x}{\phi_c}\right)^2
+\left({\partial_x}{\phi_\sigma}\right)^2\right)\\ 
+ {J_z}S^z{\partial_x}{\phi_\sigma}{\hskip -0.03cm}(0) \: + \:\, 
{J_\perp}\, \left({S^+}  e^{-i \sqrt{2}{\phi_\sigma}{\hskip -0.03cm}(0)} +
{S^-}e^{i\sqrt{2}{\phi_\sigma}{\hskip -0.03cm}(0)}\right)
\end{multline}
where $\psi_{\uparrow,\downarrow} e^{i({\phi_c}\pm{\phi_\sigma})/\sqrt{2}}$. The charge mode
${\phi_c}$ does not interact with the Kondo impurity. Dropping this
mode, we see that this is very similar to
(\ref{eqn:bosonic-p+ip}) if we identify
${J_\perp}=\lambda_\sigma$. The main difference is that
the exponential operators in (\ref{eqn:1-ch-Kondo-boson})
are $e^{\pm i \sqrt{2}{\phi_\sigma}{\hskip -0.03cm}(0)}$
rather than $e^{\pm i {\phi_\sigma}{\hskip -0.03cm}(0)/2}$, but as we will see
in the next section, this difference is unimportant.
Furthermore, the ${\partial_x}\phi_\sigma$ term in
(\ref{eqn:1-ch-Kondo-boson}) has an $S^z$,
unlike (\ref{eqn:bosonic-p+ip}).

If we follow a similar bosonization procedure in the
two-channel case, following ref. \onlinecite{Emery92},
then the electron operator is written as
\begin{equation}
\psi_{\uparrow,\downarrow;\epsilon} e^{i(\left({\phi_c}+\epsilon{\phi_f}\right) \pm
\left({\phi_\sigma}+\epsilon\phi_{\sigma f}\right))/2}
\end{equation}
where $\epsilon=\pm 1$ signify the two channels.
The Kondo Hamiltonian (\ref{eq:Kondo}) then takes the form:
\begin{multline}
{\cal H}_\text{2-ch. Kondo} =  \int dx\,\frac{v_F}{2\pi}\biggl
(\left({\partial_x}{\phi_\sigma}\right)^2
+ \left({\partial_x}{\phi_{\sigma f}}\right)^2
\biggr) \\
+ \; {J_\perp}\! \left({S^+}  e^{-i {\phi_\sigma}\!(0)} +
{S^-}e^{i{\phi_\sigma}\!(0)}\right)\!
\cos\left(\phi_{\sigma f}{\hskip -0.03cm}(0)\right)\\
+ \;  {J_z}S^z{\partial_x}{\phi_\sigma}{\hskip -0.03cm}(0) \;
\end{multline}
where we have omitted two bosons which do not couple to the Kondo impurity.
If we identify $\phi_{\sigma f}$ with $\phi_\rho$
in (\ref{eqn:bosonic-Pfaffian}), then this is very similar to the
Hamiltonian for a point contact in a $\nu=5/2$ Moore-Read Pfaffian
state. The difference between
$e^{\pm i {\phi_\sigma}{\hskip -0.03cm}(0)}$ and
$e^{\pm i {\phi_\sigma}{\hskip -0.03cm}(0)/2}$
is unimportant, as in the one-channel case.
However, the factor of 2 difference between $\cos\left(\phi_{\sigma f}\right)$
and $\cos\left(\phi_{\rho}/2\right)$ results from exchanging the
fermion field in the two-channel Kondo problem for the spin field of
our problem. This difference is important, and
will be discussed in the next section. Equation \ref{eqn:bosonic-Pfaffian}
has an extra term, the $\lambda_{1/2}$ term, which we will
also discuss in the next section.

Resonant tunneling between two semi-infinite Luttinger liquids
is also described by a very similar Hamiltonian\cite{Kane92_Resonant}.
To see this, consider first an infinite one-dimension spinless 
fermion system.   In the absence of interactions, the right and left moving
fermions,  $\psi_{R/L} \sim e^{i\varphi_{R/L}}$, are not coupled together.
With interactions present it is convenient to define new chiral boson fields,
\begin{equation}
\phi_{R/L} = \frac{1}{2} ( \sqrt{g} \pm 1/\sqrt{g} ) \varphi_R + 
\frac{1}{2} ( \sqrt{g} \mp 1/\sqrt{g} ) \varphi_L ,
\end{equation}
where the dimensionless conductance $g$ gives a measure of the interaction strength,
with $g<1$ for repulsive interactions and $g>1$ for attractive interactions.
The Hamiltonian is simply,
\begin{equation}
{\cal H}_{Lutt} = \int dx\,\frac{v}{2\pi}
\left(\left({\partial_x}{\phi_R}\right)^2
+\left({\partial_x}{\phi_L}\right)^2\right) .
\end{equation}
As before, the operator
$e^{ia\phi_{R/L}}$ 
has scaling dimension $a^2/2$.

Now consider breaking the system at $x=0$ into two semi-infinite 
Luttinger liquids, which we denote as $a,b$.  The incident  chiral bosons
$\phi_{R/L}$ are  
completely reflected at $x=0$, and so it is convenient to define two separate
chiral bosons for the two semi-infinite Luttinger liquids,
\begin{eqnarray*}
&\phi_a(x<0) \equiv \phi_R(x)  ; \hskip0.5cm &\phi_a(x>0) \equiv \phi_L(-x)  ,\\&\phi_b(x<0) \equiv \phi_L(-x)  ; \hskip0.5cm &\phi_b(x>0) \equiv \phi_R(x)  ,
\end{eqnarray*}
so that the appropriate Hamiltonian is of the same form,
\begin{equation}
{\cal H}_0 = \int dx\,\frac{v}{2\pi}
\left(\left({\partial_x}{\phi_a}\right)^2
+\left({\partial_x}{\phi_b}\right)^2\right) .
\end{equation}

Now introduce a 
quantum dot between the two semi-infinite Luttinger liquids, which is assumed to have a single state at the Fermi energy,
occupied (with $S^z = 1/2$) or unoccupied
(with $S^z=-1/2$).  Fermions on the ends of the two leads
are allowed to hop on and off the quantum dot with tunneling amplitude $t$.
The tunneling Hamiltonian is,
\begin{equation}
{\cal H}_{tun} = t ( S^+ e^{i \phi_a(x=0)/\sqrt{g}} + h.c.) + t
( S^+ e^{i \phi_b(x=0)/\sqrt{g}} + h.c.)  .
\end{equation}
Finally, upon changing variables one last time,
\begin{equation}
\phi_{a/b} = \frac{1}{\sqrt{2}} ( \phi_\sigma \pm \phi_\rho) ,
\end{equation}
one can readily see that for $g=2$ 
the full Hamiltonian ${\cal H}_0 + {\cal H}_{tun}$ is identical to 
the point contact Hamiltonian, ${\cal H}_{5/2}$ in Eq.~(\ref{eqn:bosonic-Pfaffian}), with $\lambda_{1/4}=t$ and $\lambda_{1/2} = \lambda_{\psi,0}
=0$.

\section{Strong Coupling Fixed Point and Instanton Expansion}
\label{sec:instanton}

We now analyze the Hamiltonians (\ref{eqn:bosonic-p+ip}) and
(\ref{eqn:bosonic-Pfaffian}).

\subsection{$p+ip$ superconductor with $\lambda_\psi = 0$}

When $\lambda_\psi =0$, the Hamiltonian for the 
$p+ip$ superconductor, ${\cal H}_{p+ip}$ in (\ref{eqn:bosonic-p+ip}),
has an extra symmetry, being invariant under $\phi_\sigma \rightarrow - \phi_\sigma$
together with a $\pi$ rotation of the spin about the $x-$axis:
\begin{eqnarray}
\label{eqn:K-W-sym}
{\phi_\sigma}&\rightarrow&-{\phi_\sigma}\cr
{S^z} &\rightarrow& -{S^z}\cr
S^{\pm} &\rightarrow& S^{\mp}
\end{eqnarray}
This is a Kramers-Wannier duality symmetry for the non-chiral Ising model.
This symmetry is shared by the one-channel Kondo Hamiltonian in (\ref{eqn:1-ch-Kondo-boson}).   The potentially important remaining differences
between the two Hamiltonians when $\lambda_\psi =0$ are, firstly,
an additional term of the form $J_z S^z \partial_x \phi_\sigma$  
present in the Kondo Hamiltonian, and, secondly 
the exponential operators in the Kondo problem are 
$e^{\pm i \sqrt{2}{\phi_\sigma}{\hskip -0.03cm}(0)}$
rather than $e^{\pm i {\phi_\sigma}{\hskip -0.03cm}(0)/2}$
for the $p+ip$ superconductor. However, under the 
unitary transformation generated by
\begin{equation}
\label{eqn:unitary-transf}
U=\exp(i{S^z}{\phi_\sigma}/2),
\end{equation}
these exponential factors can be readily eliminated from (\ref{eqn:bosonic-p+ip}), 
\begin{multline}
U \,{\cal H}_{p+ip}\, {U^\dagger} =\int dx\left(\frac{v_n}{2\pi}{\left({\partial_x}{\phi_\sigma}\right)^2}\right) \\
+ \lambda_{\sigma} {S^x}
- {v_n} S^z
{\partial_x}{\phi_\sigma}{\hskip -0.03cm}(0)  .
\label{eqn:Toulouse-p+ip}
\end{multline}
Similarly, under the unitary transformation, $\tilde{U} = e^{i \sqrt{2} S^z \phi_\sigma}$, we can eliminate the exponential factors from 
(\ref{eqn:1-ch-Kondo-boson}),
\begin{multline}
\tilde{U} \,{\cal H}_\text{1-ch. Kondo}\, \tilde{U^\dagger} =\int dx\left(\frac{v_F}{2\pi}{\left({\partial_x}{\phi_\sigma}\right)^2}\right) \\
+ J_{\perp} {S^x}
+ (J_z- v_F )S^z
{\partial_x}{\phi_\sigma}{\hskip -0.03cm}(0) .
\label{Toulose_Kondo}
\end{multline}
This demonstrates the equivalence of the crossover from weak to strong coupling
in the point contact
in a $p+ip$ superconductor when $\lambda_\psi=0$, to the analogous crossover in the single channel anisotropic Kondo problem with $J_z=0$.

Moreover, varying the value of $J_z$ in the anisotropic Kondo problem, while
affecting the scaling dimension of various operators in the ultraviolet, does
not effect the behavior in the infrared.  
The reason for this is that the $J_\perp S^x$ perturbation at the ultraviolet fixed point is strongly relevant, whereas the 
$(J_z-v_F)S^x \partial_x \phi_\sigma$ operator is a marginal perturbation.
Upon scaling down in energies, crossing over to the infrared fixed point,
the $J_\perp$ term rapidly grows in importance, pinning the value of the
spin to $S_x=1/2$.  Then, upon integrating out the spin degree of freedom,
the $(J_z-v_F)S^x \partial_x \phi_\sigma$ term will generate 
irrelevant terms such as $(\partial_x \phi_\sigma (0))^2$.
We mention in passing, that the soluble Toulouse limit corresponds to $J_z=v_F$,
which entirely decouples the spin from the bosonic field $\phi_\sigma$.

We can now use the well-understood behavior of the single-channel
anisotropic Kondo problem to find the infrared behavior
of the point contact in a $p+ip$ superconductor (with $\lambda_\psi=0$).
As discussed above, 
in the infrared limit the Kondo spin points along the $x$-direction,
${S^x}=1/2$, and we can perturbatively eliminate entirely the spin degree of freedom.
However, the unitary transformation
(\ref{eqn:unitary-transf}) has made ${\phi_\sigma}$ discontinuous
at $x=0$. This is the $\pi/2$ phase shift which occurs at
the strong-coupling fixed point of the single-channel Kondo problem.
A $\pm \pi/2$ phase shift for $\phi_\sigma$
corresponds in fermionic language to
\begin{equation}
\label{eqn:Majorana-bc}
{\psi_1}{\hskip -0.03cm}({0^+}) = \pm{\psi_2}{\hskip -0.03cm}({0^-})
\, , \; {\psi_2}{\hskip -0.03cm}({0^+}) = \mp{\psi_1}{\hskip -0.03cm}({0^-}).
\end{equation}
Thus, it translates, in our problem, to perfect
backscattering of the Majorana fermions at the point contact.

Hence, the RG flow from the weak-coupling fixed point for 
a point contact in a $p+ip$ superconductor (the 
Hamiltonian (\ref{eqn:bosonic-p+ip}) with $\lambda_\sigma = \lambda_\psi=0$), for non-zero $\lambda_\sigma$ crosses over 
to the strong-coupling fixed point which we discussed in section \ref{sec:point-contact}.
The leading irrelevant operator at the strong-coupling fixed point
is the dimension-$2$ operator $(\partial_x{\phi_\sigma})^2$ (which leads
to a low-temperature resistivity $\sim T^2$ in the Kondo problem).
(As discussed above, this can be obtained by integrating out the gapped fluctuations of $S^z$ about
the ground state, $\left\langle{S^z}\right\rangle=0$.) In terms of
the Majorana modes, this operator is:
\begin{equation}
\label{eqn:p+ip-leading-irrel}
(\partial_x{\phi_\sigma})^2 \sim {\psi_a}{\partial_x}{\psi_a} +
      {\psi_b}{\partial_x}{\psi_b}\ .
\end{equation}
Therefore, the leading irrelevant operator at the strong coupling fixed
point does not couple the two edges. It merely shifts their velocities locally.
The next order term obtained by integrating out the fluctuations of $S^z$
does couple the edges:
\begin{equation}
\label{eqn:p+ip-next-irrel}
({\partial_x}{\phi_\sigma})^4 \sim
\left({\psi_a}{\partial_x}{\psi_a}\right) \,\left({\psi_b}{\partial_x}{\psi_b}\right)
\end{equation}
This is the leading irrelevant operator coupling the two edges,
whose coupling constant was called $t_{\psi,\text{kin}}$ in
(\ref{eqn:strong-coupling-general}) and (\ref{eqn:strong-coupling-RG-general}).
It may be interpreted as tunneling a $p$-wave pair of neutral
fermions from one edge to the other.

\subsection{$p+ip$ superconductor with $\lambda_\psi \ne 0$}

With non-zero $\lambda_\psi$ the behavior in the infrared changes
qualitatively.  The reason for this is that the $\lambda_\psi \partial_x \phi_\sigma$ 
term does not respect the Kramers-Wannier duality symmetry, being odd under
$\phi_\sigma \rightarrow -\phi_\sigma$.  As a result, the Hamiltonian
${\cal H}_{p+ip}$ is no longer symmetry equivalent to the 
one-channel Kondo model.  The importance of this broken symmetry can be more readily appreciated after transforming the $p+ip$ Hamiltonian with the unitary transformation in (\ref{eqn:unitary-transf}), which with
non-zero $\lambda_\psi$ is now given by:   
\begin{multline}
U \,{\cal H}_{p+ip}\, {U^\dagger}= \int dx\left(\frac{v_n}{2\pi}{\left({\partial_x}{\phi_\sigma}\right)^2}\right) - \frac{\lambda_\psi}{2} S^z\\
+ \lambda_{\sigma} {S^x}
+ \left(\frac{\lambda_{\psi}}{2\pi}- {v_n} {S^z}\right)
{\partial_x}{\phi_\sigma}{\hskip -0.03cm}(0)  .
\label{p+ip_lambda_psi}
\end{multline}
Notice the presence of the $\lambda_\psi S^z$ term, which implies that in the infrared 
we can no longer drop the last term above.  Indeed, upon flowing towards the
infrared the impurity spin will no longer point along the $x-$axis, but will have a non-zero value of $S^z$:
\begin{equation}
\langle S^z  \rangle = \frac{\lambda_\psi/2}{\sqrt{\lambda_\sigma^2 + (\lambda_\psi/2)^2}}  .
\end{equation}
This implies the presence of the marginal perturbation, $\langle S^z \rangle \partial_x \phi_\sigma(0)$, at the $\lambda_\psi = 0$ strong coupling infrared fixed point.
In terms of Majorana fermions, at the strong coupling fixed point this operator corresponds to a tunneling of a Majorana fermions across the point contact,
\begin{equation}
\partial_x \phi_\sigma (0) \sim i \psi_a(0)  \psi_b(0)  .
\end{equation}
As a result, in the presence of a small non-zero $\lambda_\psi$ in the ultraviolet, the Majorana fermions in the infrared will no longer be perfectly backscattered, but will
have a small amplitude for transmission.

\subsection{$\nu=\frac{5}{2}$ QH State with $\lambda_{\psi,0}\neq 0$,
$\lambda_{1/2}=\lambda_{1/4}=0$.}
We now turn to the more complicated case of
the $\nu=5/2$ fractional quantum Hall state,
but first we consider some simpler special cases.
If $\lambda_{1/2}=\lambda_{1/4}=0$ and only
$\lambda_{\psi,0}$ is non-vanishing, then the Hamiltonian
is quadratic:
\begin{multline}
{\cal H}_{5/2}= \int dx\left(\frac{v_c}{2\pi}{\left({\partial_x}{\phi_\rho}\right)^2}
+\frac{v_n}{2\pi}{\left({\partial_x}{\phi_\sigma}\right)^2}\right)\\
\, + \,\frac{\lambda_{\psi,0}}{2\pi}\,
{\partial_x}{\phi_\sigma}{\hskip -0.03cm}(0)
\end{multline}
The $\lambda_{\psi,0}$ term can be eliminated by shifting
${\phi_\sigma}(x)~\rightarrow~{\phi_\sigma}(x) +
\frac{1}{2v_n}\lambda_{\psi,0}\,\theta(x)$.
Hence, this tunneling operator causes incoming Majorana fermions
to be scattered from one edge to the other according to:
\begin{equation}
{\psi_a}{\hskip -0.03cm}({0^+}) + i {\psi_b}{\hskip -0.03cm}({0^+}) e^{i\lambda_{\psi,0}/{2v_n}}
\left({\psi_a}{\hskip -0.03cm}({0^-}) + i {\psi_b}{\hskip -0.03cm}({0^-})\right)
\end{equation}
This should affect thermal transport, but
charge transport is completely unaffected,
i.e. $R_{xx}=0$, in this special case.

This free theory lies along a fixed line connecting the limit in
which ${\psi_a}$ and ${\psi_b}$ are unaffected by the point contact
and the other extreme in which they are switched. In appendix
\ref{sec:resonant-Majorana}, we analyze the inter-edge
resonant tunneling of Majorana fermions via a zero mode on a localized $e/4$
quasiparticle or superconducting vortex. In this case, there
is an RG flow between these two limits.

\subsection{$\nu=\frac{5}{2}$ QH State with $\lambda_{1/2}\neq 0$,
$\lambda_{\psi,0}=\lambda_{1/4}=0$.}

When $\lambda_{1/2}\neq 0$ is the only non-vanishing coupling,
the Hamiltonian takes the form:
\begin{multline}
\label{eqn:charge-half-Hamiltonian}
{\cal H}_{5/2} = \int dx\left(\frac{v_c}{2\pi}{\left({\partial_x}{\phi_\rho}\right)^2}
+\frac{v_n}{2\pi}{\left({\partial_x}{\phi_\sigma}\right)^2}\right)\\
+\:\lambda_{1/2}\, \cos\!\left({\phi_\rho}{\hskip -0.03cm}(0)\right)
\end{multline}
so the neutral Majorana mode is unaffected by the point contact.
However, charge-$e/2$ quasiparticles can tunnel between the edges.
In the infrared limit, the coupling $\lambda_{1/2}$ grows large,
according to (\ref{eqn:Pfaffian-RG}), so ${\phi_\rho}{\hskip -0.03cm}(0)$
is localized in the minima of the potential:
\begin{equation}
\label{eqn:charge-half-Dirichlet}
{\phi_\rho}{\hskip -0.03cm}(0) = (2n+1)\pi
\end{equation}
This translates to a Dirichlet boundary condition on
the non-chiral boson in the `folded' setup. An incoming
charge difference between the two edges is reversed upon
passing through the point contact, since (\ref{eqn:charge-half-Dirichlet})
means that $\left({\phi_\rho}{\hskip -0.03cm}({0^+}) +
{\phi_\rho}{\hskip -0.03cm}({0^-})\right)/2= (2n+1)\pi$.
Therefore, $R_{xx}=h/10e^2$ in this limit (see appendix \ref{sec:four-terminal}
for the definition of the four-terminal resistance).

Hence, the flow is to a `mixed' fixed point: the charged mode is
in the strong constriction limit, but the neutral mode is in the weak
constriction limit. The flow into this fixed point can be understood in
terms of instantons which take the system from one minimum
of the cosine (\ref{eqn:charge-half-Dirichlet}) to another.
Suppose that ${\phi_\rho}{\hskip -0.03cm}(0,\tau)-\pi+2\pi f{\hskip -0.03cm}(\tau)$ is a solution
to the classical equations of motion of (\ref{eqn:charge-half-Hamiltonian})
which interpolates between ${\phi_\rho}{\hskip -0.03cm}(0,-\infty) = -\pi$
and ${\phi_\rho}{\hskip -0.03cm}(0,\infty) = \pi$.
Then the classical action for a multi-instanton history,
\begin{equation}
{\phi_\rho}{\hskip -0.03cm}(0,\tau)=(2n+1)\pi+
2\pi{\sum_i} {e_i}f{\hskip -0.03cm}(\tau-{\tau_i}),
\end{equation}
with ${e_i}=\pm 1$, has a Coulomb gas
form:
\begin{equation}
\label{eqn:inst-action}
S_\text{cl} = 4\,\sum_{i,j}{e_i}{e_j}\,\ln\left|{\tau_i}-{\tau_j}\right|_{\tau_c},
\end{equation}
where the subscript on the right signifies that this
is the asymptotic form
for $\left|{\tau_i}-{\tau_j}\right|\gg{\tau_c}$, where $\tau_c$ is a short-time cutoff.
The prefactor on the right-hand-side of (\ref{eqn:inst-action})
is, more generally, given by
${\left({\Delta{\phi_\rho}{\hskip -0.03cm}(0)_\text{inst}}/{\pi}\right)^2}$,
where $\Delta{\phi_\rho}{\hskip -0.03cm}(0)_\text{inst}$ is
the separation of the minima between which the instanton interpolates.
In this case, $\Delta{\phi_\rho}{\hskip -0.03cm}(0)_\text{inst}=2\pi$.

If we sum over the possible numbers of instantons and integrate
over their (temporal) locations $\tau_i$,
we have a contribution to the partition function
\begin{equation}
\label{eqn:instanton-gas}
Z_\text{inst} = {\sum_N}\sum_{{e_m}=\pm 1}\int {\prod_{k=1}^N} {d\tau_k}
\prod_{i>j}\left|{\tau_i}-{\tau_j}\right|^{4{e_i}{e_j}}_{\tau_c}\ .
\end{equation}
We observe that this instanton gas expansion (\ref{eqn:instanton-gas})
about the strong coupling fixed point is the same as the perturbative expansion
of
$$
{\cal H}^\text{dual} = \int dx\left(\frac{v_c}{2\pi}{\left({\partial_x}{\phi_\rho}\right)^2}
+\frac{v_n}{2\pi}{\left({\partial_x}{\phi_\sigma}\right)^2}\right)
+\:t_{1}\cos\!\left(2{\phi_\rho}{\hskip -0.03cm}(0)\right).
$$
This is equivalent, in the unflipped setup, to:
\begin{multline}
S = \int d\tau\, dx\,\left({\cal L}_\text{edge}({\psi_L},{\phi_L})
+ {\cal L}_\text{edge}({\psi_R},{\phi_R})\right)\\
+ \int d\tau\, {t_1}\, \cos(({\phi_R}(0)-{\phi_L}(0))\sqrt{2})
\end{multline}
The leading irrelevant coupling at this `mixed' fixed point,
$t_1$, therefore tunnels a charge-$e$ boson across the point contact 
({\it not} an electron).
However, this is peculiar to the
special case in which $\lambda_{1/4}$ is tuned to zero.
Various properties of this mixed fixed point are discussed in Ref.\
\onlinecite{Geller05}.

\subsection{$\nu=\frac{5}{2}$ QH State with $\lambda_{1/4}\neq 0$,
$\lambda_{1/2}=\lambda_{\psi,0}=0$.}

This case is similar to the two-channel Kondo
problem, but with an important difference. After the unitary transformation,
$\lambda_{1/4}$ becomes a dimension-$7/8$ coupling
(rather than dimension-$1/2$ in the two-channel Kondo case):
\begin{multline}
U \,{\cal H}_{5/2}\, {U^\dagger}  = \int dx\left(\frac{v_c}{2\pi}{\left({\partial_x}{\phi_\rho}\right)^2}
+\frac{v_n}{2\pi}{\left({\partial_x}{\phi_\sigma}\right)^2}\right)\\
+\, \lambda_{1/4}{\hskip 0.03cm} {S^x}
\cos\!\left({\phi_\rho}{\hskip -0.03cm}(0)/2\right)
\, - v_n S^z
{\partial_x}{\phi_\sigma}{\hskip -0.03cm}(0)
\label{eqn:Toulouse-Pfaffian}
\end{multline}
While this difference in dimension is quite important for
the detailed behavior of the model, the qualitative features
of the two problems is similar. The coupling $\lambda_{1/4}$
is extremely relevant, so in the infrared limit
${\phi_\rho}(0)$ is localized in the minima of the cosine
while the spin points in the corresponding direction:
\begin{equation}
\label{eqn:charge-quarter-Dirichlet}
{S_x} = \pm 1/2\:, \;\; {\phi_\rho}{\hskip -0.03cm}(0)= \left(4n+1 \pm 1\right)\pi
\end{equation}
When ${\phi_\rho}{\hskip -0.03cm}(0)$ is localized,
the charge mode is completely reflected, as before.
The neutral Majorana fermions are also completely reflected,
according to the unitary transformation, as in the $p+ip$ case above.
Hence, the strong coupling fixed point of section \ref{sec:point-contact}
is reached in the infrared limit.

We can deduce the form of the irrelevant perturbations of
the infrared fixed point by considering instantons which
connect the minima (\ref{eqn:charge-quarter-Dirichlet}).
So long as the spin points in a fixed direction (either the $+x$ or $-x$
direction), these minima are twice as far apart as those
in (\ref{eqn:charge-half-Dirichlet}).
The minimum action instantons of this type which
contribute to charge transport have
$\Delta{\phi_\rho}{\hskip -0.03cm}(0)_\text{inst}= \pm 4\pi$,
$\Delta{\phi_\sigma}{\hskip -0.03cm}(0)_\text{inst}=0$,
$\Delta {\bf S}_\text{inst}=0$. By the arguments of the previous
subsection, these instantons correspond to the irrelevant
tunneling operator
\begin{equation}
\label{eqn:pair-hopping-pert}
H^{\rm tun}_{\rm pair} = t_\text{pair} \,
\cos(4{\phi_\rho}{\hskip -0.03cm}(0)) 
\end{equation}
which tunnels a charge-$2$ boson between the two droplets.

There is a possible complication here, namely that the spin
$\vec{S}$ can also rotate as ${\phi_\rho}{\hskip -0.03cm}(0)$ is varying.
In the special case which we are considering in this subsection,
however -- in particular, when $\lambda_{\psi,0}=0$ -- the
Hamiltonian (\ref{eqn:bosonic-Pfaffian})
is invariant under the Kramers-Wannier duality symmetry
in Eq.~(\ref{eqn:K-W-sym}).
This symmetry constrains which irrelevant perturbations of the
strong coupling fixed point can appear in the flow along this
direction from the weak-coupling fixed point, just as for the case
of the $p+ip$ superconductor. For instance,
a single electron tunneling operator, which takes the form
$\Phi_\text{el} \sim {\partial_x}{\phi_\sigma}\,e^{2i\phi_\rho}$
cannot occur because it is not invariant under this symmetry.

We can understand this in instanton language as follows
by treating the spin as a charged particle on the surface of a sphere
with a magnetic monopole at its center.
For a single electron to
tunnel, an instanton with $\Delta{\phi_\rho}=\pm 2\pi$,
$\Delta{\phi_\sigma}=0$ is needed. Such an instanton only
connects two minima of the Hamiltonian if the spin
$S$ is also reversed, e.g. from ${S_x}=1/2$ to
${S_x}=-1/2$. There are many such equally good classical paths
from one point on the sphere to its antipode, but they
will contribute with different Berry phases, as result of
the monopole, and cancel.

Therefore, the dimension-$8$ operator (\ref{eqn:pair-hopping-pert})
is the leading irrelevant
operator in the infrared when $\lambda_{1/4}$ is the only non-zero relevant
perturbation in the ultraviolet. Hence, we obtain the non-generic
low-temperature transport of (\ref{eqn:low-T-non-generic}).

\subsection{$\nu=\frac{5}{2}$ QH State, General Case:
$\lambda_{1/4}$, $\lambda_{1/2}$, $\lambda_{\psi,0}\neq 0$}

The approach to the limit of two decoupled droplets
is so rapid in the previous special case
because a single electron can't tunnel
from left to right. 
However, when $\lambda_{\psi,0}$ is also non-zero, the
Hamiltonian (\ref{eqn:bosonic-Pfaffian}) is no longer invariant
under the symmetry (\ref{eqn:K-W-sym}). Therefore, an electron
tunneling term is not forbidden.
Such a term arises from an instanton gas expansion
because the term
$ i \lambda_{\psi,0} {\psi_1} {\psi_2} = \lambda_{\psi,0}{\partial_x}\phi_\sigma$
leads to a term $\lambda_{\psi,0} S^z$ after application
of the unitary transformation (\ref{eqn:unitary-transf}), just as for the $p+ip$ superconductor as seen explicitly in Eq.~(\ref{p+ip_lambda_psi}). The symmetry between
the different classical paths is now broken, and there
is a unique minimum action instanton in spin space
connecting two classical minima such as ${S_x} = 1/2$,
${\phi_\rho}(0)= 2\pi$ and ${S_x} = -1/2$,
${\phi_\rho}(0)= 0$. This instanton transfers charge $e$
while simultaneously flipping the sign of the neutral
part of the quasiparticle tunneling operator ${\sigma_a}{\sigma_b}$.
(Thereby leaving the Hamiltonian unchanged.)
Therefore, it corresponds to the electron tunneling operator
\begin{eqnarray}
\nonumber
H^{\rm tun}_{\rm el} &=&  t_\text{el} \,{\partial_x}{\phi_\sigma}{\hskip -0.03cm}(0)\,
\cos(2{\phi_\rho}{\hskip -0.03cm}(0))\\
&=&  t_\text{el}\, i{\psi_a}{\hskip -0.03cm}(0){\psi_b}{\hskip -0.03cm}(0)\,
\cos(2{\phi_\rho}{\hskip -0.03cm}(0))
\end{eqnarray}
The presence of non-zero $\lambda_{1/2}$ does not lift the
degeneracy between the minima;
it just makes them deeper and suppresses the maxima.
Therefore, it does not change the analysis above.
Hence, we now obtain the generic low-temperature
resistance of (\ref{eqn:low-T-non-generic}).

\section{Discussion}

We have developed a framework to describe quantitatively the tunneling
of edge modes in two-dimensional systems with non-Abelian statistics.
Since the edge modes are both chiral and have non-trivial fusion
rules, we utilized the formalism of Moore and Seiberg to first define
the tunneling operator uniquely, and then compute its
conformal blocks.

One immediate result of our mapping onto the Kondo problem is the
entropy loss in the flow from the ultraviolet to the infrared.  In
Ref.\ \onlinecite{Fendley06c}, we discuss in depth the entanglement
entropy loss in such systems. We define and compute {\em
holographic partition functions}, which describe the entanglement
entropy of both bulk and edge quasiparticles in conformal field theory
language.

Unfortunately, experiments on fractional quantum Hall systems do not
generally measure the entropy or the specific heat, but instead
transport quantities such as the tunneling current, which
we have discussed here. To compute transport quantities
in the non-Abelian case, one must utilize the precise definition
of the tunneling operator which we have given.
Moreover, we showed that the perturbative expansion is given in
terms of conformal blocks, not the usual non-chiral correlation
functions. Transport noise measurements may also
shed light on the properties of non-Abelian quasiparticles \cite{Bena06}.
The Keldysh formalism, useful for non-equilibrium transport
situations, will require some refinement to be
used in such chiral theories, building on the formulation
of tunneling given here.

We have mainly focused on the simplest type of tunneling, which does
not change the fusion channel of the edge modes-- no qubit is flipped
as a result of tunneling. This is the only possibility for tunneling
through a simple point contact, but more complicated geometries can
result in more complicated tunneling events. To give the simplest of
such possibilities, consider an antidot at the point contact. Then a
tunneling event can cause a fermion to leave or join the edge of the
antidot, effectively adding or removing a zero mode from the bulk of
the system. Such a tunneling event would correspond to the $\sigma$
quasiparticle annihilation and creation operators on the two edges to
be in the $\psi$ channel, instead of the $I$ channel. More complicated
geometries, such as an antidot
with two point contacts at both ends, can result in more complicated
possibilities. A single tunneling event is described by
two-dimensional vector space of operators, so for a given geometry,
the tunneling operator can be any element of this vector space.

Our formalism should be applicable to any gapped two-dimensional
quantum system with gapless chiral edge modes. Moore and Seiberg's
results (and hence ours) apply to any rational conformal field theory;
``Rational'' means that there are a finite
number of chiral primary fields under some extended symmetry
algebra. In the fractional quantum Hall effect, the extended symmetry
algebra of the edge modes arises from the requirement that all
quasiparticle and quasihole creation/annihilation operators commute
with the electron operator. This effectively turns the electron
operator into a symmetry operator\cite{Moore91},
making it likely that the resulting
edge conformal field theory is rational. 

One obvious candidate for applying our results is the Read-Rezayi
states\cite{Read99}, which extend the Moore-Read state to an entire
series of fractional quantum Hall states with non-Abelian
braiding. Here, the description of the edge modes in terms of rational
conformal field theory is already known \cite{Read99}, so defining the
appropriate conformal blocks is straightforward. Bosonizing these
conformal blocks is not straightforward, but may be
possible. Bosonizing the Moore-Read tunneling operator was possible
because the edge modes are described by the chiral Ising model, which
has central charge $c=1/2$; since there are two edges, the combined
theory has $c=1$, the central charge of the free boson.  The
non-Abelian part of the $k$th Read-Rezayi theory is the $Z_k$
parafermion conformal field theory, which has central charge
$c=2(k-1)/(k+2)$. To bosonize such theories, one must combine them
with other theories to get $c$ integer. This was done for the $Z_3$
chiral theory (or to be precise, for the equivalent 1+1 dimensional
quantum impurity problem) in Ref.\ \onlinecite{Affleck01}. We expect
that this analysis can be adapted to our situation.

We also note that
while this formalism is necessary to define precisely the 
tunneling
for non-Abelian states, we
believe it may be fruitful to utilize it even for abelian
fractional quantum Hall states. For more general states than
Laughlin's, the structure of quasiparticles and the phases which
result under their braiding can get quite elaborate.  Whereas keeping
track of phases is tricky in any formalism, exploiting the symmetry
algebra may provide a useful tool for simplifying the analysis.

\acknowledgements
We would like to thank E.\ Fradkin, M.\ Freedman, E.-A.\ Kim, A.\ Kitaev, 
A.\ Ludwig, J.\ Preskill, N.\ Read, and A.\ Stern for discussions. 
This research has been supported by the NSF under grants  
DMR-0412956 (P.F.), PHY-9907949 and DMR-0529399 (M.P.A.F.) and 
DMR-0411800 (C.N.), and by the ARO under grant W911NF-04-1-0236 (C.N.).

\appendix

\section{Four-Terminal Transport at $\nu=5/2$}
\label{sec:four-terminal} 
 
\begin{figure}[t!]
\centerline{\includegraphics[width=3.25in]{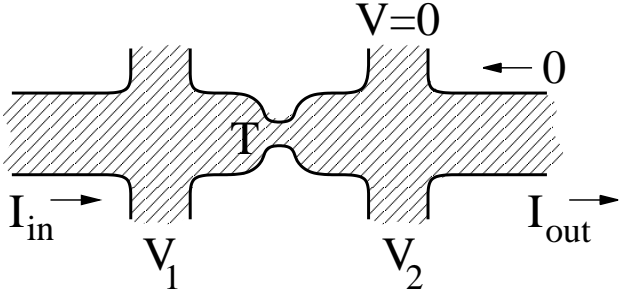}}
\caption{Four-terminal transport setup. Voltages
$V_1$ and $V_2$ are measured at the lower left and right,
respectively. Current $I_\text{in}$ is injected and
current $I_\text{out}$ flows out. When there is no
inter-edge tunneling between the top and bottom edges, the transmission coefficient vanishes,
$T=0$. In the strong constriction limit, $T=1$.}
\label{fig:4-terminal}
\end{figure}

In this appendix, we discuss four-terminal transport
in the $\nu=5/2$ quantum Hall state in the presence of
a point contact. In figure \ref{fig:4-terminal}, an
current $I_\text{in}$ is injected along the lower edge
at the left. The voltage to the left of the point contact
is related to this current by the Hall relation:
\begin{equation}
I_\text{in} = \frac{5}{2}\,\frac{e^2}{h}\,{V_1}
\end{equation}
For notational convenience, we set the voltage
of the top edge at the right of the point contact
to zero. Then the current going out to the right
is related to the voltages on the bottom edge
to the right and left of the point contact according to:
\begin{equation}
I_\text{out} = \frac{5}{2}\,\frac{e^2}{h}\,{V_2} = 
\left[2+\frac{1}{2}(1-T)\right]\frac{e^2}{h}\,{V_1}
\end{equation}
where $T$ is the transmission coefficient
for tunneling between the edges. Hence,
\begin{eqnarray}
R_{xx} \equiv \frac{{V_1}-{V_2}}{I_\text{out}} &=& \frac{1}{\left[2+\frac{1}{2}(1-T)\right]\frac{e^2}{h}}
- \frac{1}{\frac{5}{2}\,\frac{e^2}{h}}\cr
&=& \frac{h}{e^2}\,\frac{2}{5}\,\frac{(T/5)}{1-\frac{T}{5}}
\end{eqnarray} 
In the strong coupling limit, $T=1$, and $R_{xx}=\frac{h}{e^2}\,\frac{1}{10}$.
At small, non-zero temperature, $1-T \sim {t_\text{el}^2}\,{T^4}$,
so we find:
\begin{equation}
R_{xx} - \frac{1}{10}\,\frac{h}{e^2} \sim - {t_\text{el}^2}\,{T^4}
\end{equation}

\section{Explicit Computation of Conformal Blocks} 
\label{sec:block-computation}

Ordinarily, one is not interested in the explicit forms of the conformal blocks
of the $2n$-point spin-field correlation functions in the Ising model.
They are merely an intermediate step on the way to computing the
quantities which are actually of interest, the non-chiral correlation functions.
The latter can be computed by the methods of ref. \onlinecite{DSZboso}.
However, in the context of topological states and their edge excitations,
the explicit forms of conformal blocks themselves are important quantities.
Here, we show through some examples how they can be obtained using
the methods of section \ref{sec:bosonization}.

{}From (\ref{eqn:blocks-bosonized}), we see, for instance
that:
\begin{multline}
(0,0)^2 = \Bigl\langle
\Big(e^{i\left({\phi_\sigma}({z_1})
-{\phi_\sigma}({z_2})\right)/2}
+ e^{-i\left({\phi_\sigma}({z_1})
-{\phi_\sigma}({z_2})\right)/2} \Big)\\
\times\,\Big(e^{i\left({\phi_\sigma}({z_3})
-{\phi_\sigma}({z_4})\right)/2}
+ e^{-i\left({\phi_\sigma}({z_3})
-{\phi_\sigma}({z_4})\right)/2} \Big)
\Bigr\rangle\\
= 2\Bigl\langle e^{\frac{i}{2}{\phi_\sigma}{\hskip -0.03cm}({z_1})}
e^{-\frac{i}{2}{\phi_\sigma}{\hskip -0.03cm}({z_2})}
e^{\frac{i}{2}{\phi_\sigma}{\hskip -0.03cm}({z_3})}
e^{-\frac{i}{2}{\phi_\sigma}{\hskip -0.03cm}({z_4})}
\Bigr\rangle\\
+ 2\Bigl\langle e^{\frac{i}{2}{\phi_\sigma}{\hskip -0.03cm}({z_1})}
e^{-\frac{i}{2}{\phi_\sigma}{\hskip -0.03cm}({z_2})}
e^{-\frac{i}{2}{\phi_\sigma}{\hskip -0.03cm}({z_3})}
e^{\frac{i}{2}{\phi_\sigma}{\hskip -0.03cm}({z_4})}
\Bigr\rangle
\end{multline}
In obtaining the second equality, we used the symmetry
of the free boson theory under ${\phi_\sigma}\rightarrow-{\phi_\sigma}$.
In the same way, we can obtain $(1,1)^2$.

Hence, we have:
\begin{eqnarray*}
(0,0) &=& \left[ \left(\frac{z_{13} z_{24}}{z_{12} z_{14} z_{23} z_{34}}\right)^{1/4}
+ \left(\frac{z_{14} z_{23}}{z_{12} z_{13} z_{24} z_{34}}\right)^{1/4}\right]^{1/2}\cr
(1,1) &=& \left[ \left(\frac{z_{13} z_{24}}{z_{12} z_{14} z_{23} z_{34}}\right)^{1/4}
- \left(\frac{z_{14} z_{23}}{z_{12} z_{13} z_{24} z_{34}}\right)^{1/4}\right]^{1/2}
\end{eqnarray*}
These expressions are the same as ${\cal F}_I$, ${\cal F}_\psi$
in (\ref{fourpointexplicit}), which were obtained by solving
a differential equation following from the existence
of null vectors in ref. \onlinecite{Belavin84}. Higher-point
conformal blocks can be obtained by solving even more complicated
differential equations. However, we can obtain all of these
by similar bosonization formulas as above. For instance,
a calculation of a conformal block of a six-point function
gives:
\begin{multline}
(0,0,0) \Biggl[ \left(\frac{z_{13} z_{15} z_{35} z_{24} z_{26} z_{46}}{z_{12} z_{14} z_{16} z_{23} z_{34} z_{36} z_{25} z_{45} z_{56}}\right)^{1/4}\\
+\: \left(\frac{z_{13} z_{16} z_{36} z_{24} z_{25} z_{45}}{z_{12} z_{14} z_{15} z_{23}
z_{34} z_{35} z_{26} z_{46} z_{56}}\right)^{1/4}\\
+\: \left(\frac{z_{14} z_{15} z_{45} z_{23} z_{26} z_{36}}{z_{12} z_{13} z_{16} z_{24}
z_{34} z_{46} z_{25} z_{35} z_{56}}\right)^{1/4}\\
+\: \left(\frac{z_{13} z_{16} z_{45} z_{23} z_{25} z_{35}}{z_{12} z_{13} z_{15} z_{24}
z_{34} z_{45} z_{26} z_{35} z_{56}}\right)^{1/4}
\Biggr]^{1/2}
\end{multline}
Conformal blocks of correlation functions of $\sigma$'s and $\psi$'s
can also be obtained in this way, using (\ref{eqn:current-bosonized}).

\section{Resonant Tunneling Between Edges}
\label{sec:resonant-Majorana}

In this appendix, we discuss the situation in which
a charge-$e/4$ quasiparticle is localized in the middle
of a point contact in a Pfaffian quantum Hall
state at $\nu=5/2$. Such a quasiparticle has a
zero-energy Majorana zero mode localized at its core,
so a Majorana fermion at the edge can tunnel resonantly
through this mode to the other edge. Let us suppose that all other
types of tunneling are much smaller. (With a charged
quasiparticle localized in the point contact, Coulomb blockade
might strongly suppress
the tunneling of charge $e/4$ and $e/2$ quasiparticles.)
The action for this situation is:
\begin{multline}
S = \int dx\,d\tau\, \left({\cal L}_\text{fermion}\left({\psi_L}\right)
+ {\cal L}_\text{fermion}\left({\psi_R}\right)\right)\\
+ \psi_\text{loc}{\partial_\tau}\psi_\text{loc} +
+ i {t_R} \psi_\text{loc}{\psi_R}{\hskip -0.03cm}(0) + 
i {t_L} \psi_\text{loc}{\psi_L}{\hskip -0.03cm}(0)
\end{multline}
where $\psi_\text{loc}$ is Majorana zero mode
at the localized quasiparticle and ${t_R}$, ${t_L}$
are the hopping matrix elements between, respectively,
the right and left edges and the localized zero mode.
As we will see momentarily, a resonance occurs when
${t_R} = \pm {t_L}$ (the relative sign can be absorbed in
$\psi_\text{loc}$). Since this action is quadratic, we can diagonalize
it explicitly. We find:
\begin{eqnarray*}
\label{eqn:resonant-Majorana-scattering}
\begin{pmatrix} 
{\psi_R}{\hskip -0.03cm}({0^+},\omega)\\
{\psi_L}{\hskip -0.03cm}({0^+},\omega)
\end{pmatrix}
&=& \frac{1}{\omega + \frac{i}{2}\left({t_R^2}+{t_L^2}\right)}{\cal M}
\begin{pmatrix} 
{\psi_R}{\hskip -0.03cm}({0^-},\omega)\\
{\psi_L}{\hskip -0.03cm}({0^-},\omega)
\end{pmatrix}
\\
\nonumber {\cal M}&\equiv&
\begin{pmatrix}
\omega + \frac{i}{2}\left({t_L^2}-{t_R^2}\right) & -i\,{t_R} {t_L} \\
-i\,{t_R} {t_L} & \omega + \frac{i}{2}\left({t_R^2}-{t_L^2}\right)
\end{pmatrix}
\end{eqnarray*} 
When ${t_R} = {t_L}$, we find that
${\psi_R}{\hskip -0.03cm}({0^+},0)~=~{\psi_L}{\hskip -0.03cm}({0^-},0)$
and
${\psi_L}{\hskip -0.03cm}({0^+},0)~=~{\psi_R}{\hskip -0.03cm}({0^-},0)$.
Thus, the `RG flow' is to the strong constriction limit.

It is instructive to re-express this result in bosonic terms,
by adopting the same definition as in Eq.(\ref{bosonization-def}),
$e^{i\phi_\sigma} \sim \psi_R + i \psi_L$.  This gives,
\begin{equation}
\phi_\sigma(x=0^+) = (\pi/2) - \phi_\sigma(x=0^-)  .
\end{equation}
In striking contrast to the strong coupling $p+ip$ and Kondo fixed points
which have a simple $\pi/2$ phase shift, $\phi_\sigma(x=0^+) = \phi_\sigma(x=0^-) + (\pi/2)$, the strong coupling resonant tunneling fixed point
is qualitatively different involving not only a phase shift but
a sign change of the bosonic field.  This demonstrates that despite
the common ultraviolet fixed points in these two situations,
the different form of the two inter-edge tunneling perturbations (ie. tunneling of $\sigma$ particles versus  Majorana tunneling through a localized $\sigma$ particle, respectively) causes both the nature of the crossovers and the ``destination"
strong coupling infrared fixed points to be qualitatively distinct.
Nevertheless, the leading irrelevant operator at the two
infrared fixed points is the same, given in
(\ref{eqn:p+ip-leading-irrel}).  To see this, we linearize for small
$\omega$ in (\ref{eqn:resonant-Majorana-scattering}) to obtain
$(i\omega + t_R t_L \partial_x) (\psi_R+\psi_L)|_{x=0} = 0$.
Upon Fourier transforming back to real time  we obtain,
\begin{equation}
\partial_t (\psi_R + \psi_L) = -i [\psi_R + \psi_L, {\cal H}_{irr} ]
= \delta(x) t_R t_L \partial_x(\psi_R + \psi_L)  .
\end{equation}
This is consistent with a leading irrelevant operator of the form,
\begin{equation}
{\cal H}_{irr} \sim (i \psi_R \partial_x \psi_R + i \psi_L \partial_x \psi_L)|_{x=0}  \sim \Big( \partial_x \phi_\sigma(0) \Big)^2  ,
\end{equation}
the same as in (\ref{eqn:p+ip-leading-irrel}).

\vskip -0.5cm 
 
 

\end{document}